\documentclass[aps, prd,preprint,superscriptaddress,showpacs,amsmath,amssymb]{revtex4-1}
\usepackage[dvipdfm]{graphicx}
\usepackage{dcolumn}
\usepackage{bm}
\usepackage{multirow}

\begin{document}


\title{Observation of thundercloud-related gamma rays and neutrons in Tibet}

\author{H. Tsuchiya}
\altaffiliation{Japan Atomic Energy Agency, Tokai-mura, Naka, Ibaraki 319-1195, Japan}
\affiliation{High-energy Astrophysics Laboratory,  Riken, 2-1, Hirosawa, Wako, Saitama 351-0198, Japan}
\author{K. Hibino}
\affiliation{Faculty of Engineering, Kanagawa University, Yokohama 221-8686, Japan}
\author{K. Kawata}
\affiliation{Institute for Cosmic Ray Research, University of Tokyo, Kashiwa 277-8582, Japan}
\author{N. Hotta}
\affiliation{Faculty of Education, Utsunomiya University, Utsunomiya 321-8505, Japan}
\author{N. Tateyama}
\affiliation{Faculty of Engineering, Kanagawa University, Yokohama 221-8686, Japan}
\author{M. Ohnishi}
\affiliation{Institute for Cosmic Ray Research, University of Tokyo, Kashiwa 277-8582, Japan}
\author{M. Takita}
\affiliation{Institute for Cosmic Ray Research, University of Tokyo, Kashiwa 277-8582, Japan}
\author{D. Chen}
\affiliation{Institute for Cosmic Ray Research, University of Tokyo, Kashiwa 277-8582, Japan}
\author{J. Huang}
\affiliation{Key Laboratory of Particle Astrophysics, Institute of High Energy Physics, Chinese Academy of Sciences, Beijing 100049, China}
\author{M. Miyasaka}
\affiliation{Caltech, Pasadena, CA 91125, USA}
\author{I. Kondo}
\affiliation{High-energy Astrophysics Laboratory,  Riken, 2-1, Hirosawa, Wako, Saitama 351-0198, Japan}
\author{E. Takahashi}
\affiliation{High-energy Astrophysics Laboratory,  Riken, 2-1, Hirosawa, Wako, Saitama 351-0198, Japan}
\author{S. Shimoda}
\affiliation{High-energy Astrophysics Laboratory,  Riken, 2-1, Hirosawa, Wako, Saitama 351-0198, Japan}
\author{Y. Yamada}
\affiliation{High-energy Astrophysics Laboratory,  Riken, 2-1, Hirosawa, Wako, Saitama 351-0198, Japan}
\author{H. Lu}
\affiliation{Key Laboratory of Particle Astrophysics, Institute of High Energy Physics, Chinese Academy of Sciences, Beijing 100049, China}
\author{J.~L. Zhang}
\affiliation{Key Laboratory of Particle Astrophysics, Institute of High Energy Physics, Chinese Academy of Sciences, Beijing 100049, China}
\author{X.~X. Yu}
\affiliation{Key Laboratory of Particle Astrophysics, Institute of High Energy Physics, Chinese Academy of Sciences, Beijing 100049, China}
\author{Y.~H. Tan}
\affiliation{Key Laboratory of Particle Astrophysics, Institute of High Energy Physics, Chinese Academy of Sciences, Beijing 100049, China}
\author{S.~M. Nie}
\affiliation{Guizhou University for Nationalities, Guiyang 550025, China}
\author{K. Munakata}
\affiliation{Department of Physics, Shinshu University, Matsumoto 390-8621, Japan}
\author{T. Enoto}
\affiliation{High-energy Astrophysics Laboratory,  Riken, 2-1, Hirosawa, Wako, Saitama 351-0198, Japan}
\affiliation{Kavli Institute for Particle Astrophysics and Cosmology, Department of Physics and SLAC National Accelerator Laboratory, Stanford University, Stanford, CA 94305, USA}
\author{K. Makishima}
\affiliation{Department of Physics, University of Tokyo, 7-3-1, Hongo, Bunkyo-ku, Tokyo 113-0033, Japan}

\date{\today}
\begin{abstract}
During the 2010 rainy season in Yangbajing (4300 m above sea level) in Tibet, China,
a long-duration count enhancement associated with thunderclouds was detected by a solar neutron telescope 
and neutron monitors installed at the Yangbajing Comic Ray Observatory. 
The event, lasting for $\sim$40 min, was observed on July 22, 2010. 
The solar neutron telescope detected significant $\gamma$-ray signals with energies $>$40 MeV in the event. 
Such a prolonged high-energy event has never been observed in association with thunderclouds, 
clearly suggesting that electron acceleration lasts for 40 min in thunderclouds.
In addition, Monte Carlo simulations showed that $>$10-MeV $\gamma$ rays largely contribute to the neutron monitor signals, while
$>$1-keV neutrons produced via a photonuclear reaction contribute relatively less to the signals. 
This result suggests that enhancements of neutron monitors during thunderstorms are not necessarily a
clear evidence for neutron production, as previously thought.  
\end{abstract}

\maketitle

\section{Introduction\label{sec:intro}}
Recent observations have shown that thunderclouds are powerful particle accelerators,
emitting bremsstrahlung $\gamma$ rays that extend to 10 MeV or 
higher~\cite{growth_2007,norikura_tsuchiya_2009,fuji_torii_2009,armenia_2010,armenia_2011,monjyu_torii_2011,growth_2011}.
Unlike terrestrial gamma ray flashes (TGFs) and lightning-related $\gamma$ rays, which last for milliseconds or less, 
thundercloud-related $\gamma$ rays are characterized by durations of a few tens of seconds to a few minutes, or occasionally more than 10 min.
These thundercloud-related $\gamma$ rays have been thought to be produced by 
relativistic electrons, in accordance with the relativistic runaway electron avalanche (RREA) model~\cite{Gurevich_1992,GM_1999,Dwyer_RF_2003}
that involves acceleration and multiplication of ambient electrons. 
However, mainly because of the lack of a large sample of thundercloud-related $\gamma$ rays, 
there is still no consensus that all of those $\gamma$ rays are really generated by the RREA mechanism.
It is also unclear whether the charging mechanism of thunderclouds is related to the production of thundercloud-related $\gamma$ rays.

Several groups conducting
their experiments on high mountains have reported detecting 
various particles besides $\gamma$ rays, in possible association with thunderstorms~\citep{Alex_2002,norikura_muraki_2004}. 
Among such particles, the production of neutrons in coincidence with natural lightning
by a thermonuclear reaction $\mathrm{{}^2H + {}^2H \rightarrow  n\, (2.45 MeV\,) + {}^3He}$
was closely investigated in the 1970s$-$1980s because such neutrons provide
a key not only to elucidate the mechanism of lightning but also to know if such neutrons
are another source captured by $^{14}$C. Investigating such neutrons, 
\citet{Shah_1985} and \citet{SK_1999} reported detections of $10^7$$-$$10^{10}$ neutrons per lightning strike. 
 
Instead of the above fusion mechanism, \citet{BR_2007_Npro} proposed a photonuclear
reaction, $^{14}$N($\gamma, n){}^{13}$N, 
showing that the fusion mechanism is not feasible under the usual physical conditions in lightning.
The photonuclear reaction begins at a $\gamma$-ray energy 
of 10.5 MeV~\cite{photonuclear_1975}, and hence may occur because $\gamma$ rays with energies above 
the threshold have been actually observed. Therefore, photonuclear neutrons provide another clue to solve 
non-thermal mechanism in thunderstorms. Actually, \citet{Carlson_Npro_2010} 
made a close investigation on neutron production in TGFs, predicting  that a TGF averagely produces $\sim$$10^{12}$ neutrons
corresponding to a ground-level neutron fluence of $(0.03-1)$$\times10^{4}$ $\mathrm{m^{-2}}$. Similarly, 
\citet{Babich_prediction_gn_2010} also predict that neutrons with fluence of $10^3$$-$$10^7$ $\mathrm{m^{-2}}$
would arrive at ground level when energetic $\gamma$ rays are produced under the RREA mechanism.
Many neutron monitors, installed
at cosmic-ray stations in the world, could detect such neutron bursts in thunderstorms, if those neutrons 
actually reaching them.  

Interestingly, a clear enhancement during thunderstorms was recently detected 
by neutron monitors installed at Mt. Aragats at an altitude of 3250 m a.s.l. in Armenia~\cite{armenia_2010,armenia_2011}. In addition, 
plastic scintillator-based detectors, arranged close to the neutron monitors,  
detected long-duration (10$-$20 min) $\gamma$ rays extending to 40$-$50 MeV.
Generally, owing to its detection method, a neutron monitor is believed to be 
very sensitive to nucleons but insensitive to $\gamma$ rays and electrons. Thus, \citet{armenia_2010} have concluded that 
the observed increase of the Aragats neutron monitor is attributable to neutrons generated
via the photonuclear reaction. 

Similar to the Armenia case,  clear enhancements were occasionally obtained by
some detectors installed at the Yangbajing Cosmic Ray Observatory ($30.103^\circ$N, $90.523^\circ$E; 
cutoff rigidity $=14$ GV), which is located on a mountain 4.3 km a.s.l.~in Tibet, China. Actually, 
because two electric-field mills (BOLTEK EFM-100) were installed in February 2010 at the observatory, 
five large count enhancements were found to be associated with electric-field variations in the rainy season. 
In this paper we present one prolonged count increase with duration of 40 min,
which obtained by both the Yangbajing neutron monitor~(YBJ NM) and a solar neutron telescope~(SNT). 
Utilizing the event, we especially discuss how the observed signals are attributed to $\gamma$ rays and neutrons produced via the
photonuclear reaction. Then, we deduce fluxes of $\gamma$ rays and neutrons, and compare them with those 
obtained from other experiments and Monte Carlo predictions. 
\section{Experiment\label{sec:exp}}
Due to its high altitude (4300 m a.s.l.) and meteorological conditions from May to October,  
the sky above the Yangbajing Cosmic Ray Observatory is frequently covered with thunderclouds. The observatory has three independent 
detectors:  the Tibet air shower array~\cite{TibetIII_2009}, YBJ NM~\cite{YBJ_NM2007}, and SNT~\cite{TibetSNT}. 
The air shower array, working successfully since 1990,  mainly observes $10^{12}$$-$$10^{17}$ eV primary cosmic rays. 
On the other hand, YBJ NM and SNT have been operating since 1998, mainly aiming at detecting $>$100 MeV solar-flare neutrons 
and protons to elucidate the ion-acceleration mechanism in solar flares. 
YBJ NM and SNT are placed close to each other in one building.

\subsection{Yangbajing neutron monitor}
YBJ NM consists of 28 NM64-type detectors~\cite{NM64_1,NM64_2}
having the largest area of 32 $\mathrm{m^2}$ among world-wide 
neutron monitors. 
An NM64 neutron monitor is composed of a $\mathrm{BF_{3}}$ counter, which is surrounded by polyethylene [$(\mathrm{C}_{2}\mathrm{H}_{4})_{\mathrm{n}}$] plates of
thickness 7.5 cm and lead blocks with an average thickness of 120 $\mathrm{g\, cm^{-2}}$.
The polyethylene plates reflect low-energy nucleons accidentally produced in
substances close to the detector, while the lead blocks multiply impinging neutrons via inelastic scattering
processes. 

Each $\mathrm{BF_{3}}$ counter that contains the $\mathrm{BF_{3}}$ gas with the density of $3 \times 10^{-4}$~$\mathrm{g\,cm^{-3}}$
has a length of 190.8 cm and radius of 7.4 cm.
The counter can easily detect a thermal neutron via a neutron capture reaction 
as ${}^{10}\mathrm{B} + {}^{1}\mathrm{n} \rightarrow {}^{4}\mathrm{He} + {}^{7}\mathrm{Li}$, because
the cross section of the capture reaction increases rapidly as the kinetic energy of the neutron 
decreases to thermal energy. 
To efficiently decelerate neutrons to thermal energy by elastic collisions with hydrogen nuclei, 
each $\mathrm{BF_{3}}$ counter is inserted into an additional polyethylene tube with a thickness of 2 cm. 
A ${}^{4}\mathrm{He}$ ion created by a neutron capture reaction produces a large 
amount of ionization loss by $\sim$1 MeV or higher
in the $\mathrm{BF_{3}}$ counters to provide a sufficiently large signal on its anode. 
Due to the multiplication and thermalization of the incident neutron, the large signal has
no information about the incident energy. However, the signal can be easily distinguished from 
charged secondary cosmic-ray background events (mainly muons), which provide
a small signal of $\sim$9 keV. Output signals from individual counters are fed to the data acquisition system, 
and the event number of individual counters is recorded every second. 

It is widely believed that neutron monitors have no sensitivity to electromagnetic components because
of the thick lead blocks. However,
a photonuclear reaction between $\gamma$ rays and lead nuclei begins at the $\gamma$-ray energy of  7 MeV, and peaking at $\sim$13 MeV~\cite{photonuclear_1975}.
Thus,  high-energy $\gamma$ rays associated with energies $>$7 MeV can produce neutrons via the photonuclear reaction.
Accordingly, neutron monitors might capture such photonuclear neutrons produced by thundercloud-related
$\gamma$ rays extending to 10 MeV or higher energies. 

In order to investigate this possibility, we performed a Monte Carlo simulation based on GEANT4~\cite{GEANT4}
and derived detection efficiencies of an NM64 neutron monitor, including YBJ NM, for 
neutrons, $\gamma$ rays, electrons, and positrons in a wide energy range of 1 keV$-$1 GeV.
For this purpose, a geometry of a standard NM64 neutron monitor~\cite{NM64_2} was constructed, and $10^5$
mono-energetic particles for each species were illuminated on the same area as the neutron monitor. 
In one mono-energetic simulation, an irradiated particle was 
isotropically injected toward the neutron monitor from the vertical direction to 60 degrees.
We choose in each simulation (including air propagation simulations described later) a hadronic model of QGSP\_BERT\_HP provided by GEANT4 to treat physics processes of neutrons in the atmosphere.

\begin{figure}
\includegraphics[width=0.8\textwidth]{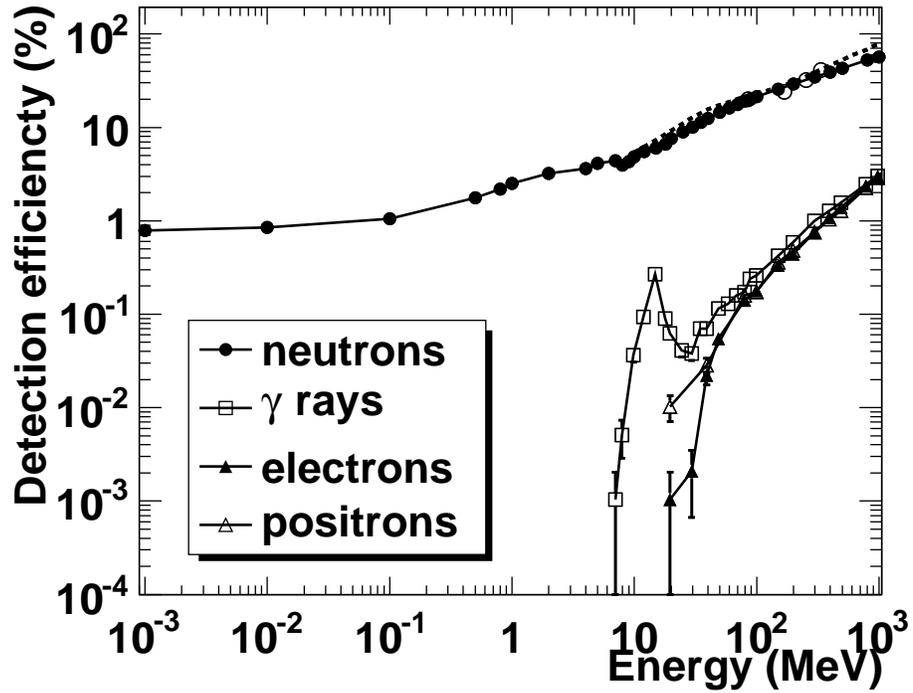}
\caption{%
Detection efficiencies of an NM64 neutron monitor for neutrons, $\gamma$ rays,
electrons, and positrons, as determined by the GEANT4 simulation. 
A dashed curve indicates detection efficiency of an NM64 neutron monitor obtained
by \citet{Clem_2000}. Open circles represent experimental results by \citet{Shibata_NM2001}.
Horizontal axis denotes incident energy in MeV. Errors are statistical 1$\sigma$ errors.
}
\label{fig:NM_eff}
\end{figure}

Figure~\ref{fig:NM_eff} shows detection efficiencies determined in this manner for the four particles. 
The present efficiency for neutrons (black circles) agrees well with that obtained by another detector simulation
conducted by \citet{Clem_2000} (dashed lines). The difference in efficiencies at 10 MeV$-$1 GeV of the two simulations 
is maximum 30\%.
In addition, our results for neutrons can reproduce well efficiencies
experimentally determined using an accelerator neutron beam~\cite{Shibata_NM2001}. These consistencies 
validate our simulation results.

As expected, the present simulation reveals that an NM64 neutron monitor 
has sensitivity to electromagnetic components in energies range 10 MeV$-$1 GeV. 
Compared with the efficiencies for neutrons, those for $\gamma$ rays in the energy
range are lower by a factor of $1/125$$-$$1/20$. 
Similarly, high-energy electrons and positrons entering the lead blocks
emit $\gamma$ rays via bremsstrahlung, which in turn generate either neutrons via the photonuclear process
or electrons via pair creation. Since the critical energy of electrons in lead is 7 MeV, these cascading processes would 
continue until energies of electrons and $\gamma$ rays are below the critical energy.
As the incident energy of electromagnetic components increases, the cascading becomes more effective in causing the photonuclear reaction. Thus, 
detection efficiencies for electromagnetic components increase (Fig.~\ref{fig:NM_eff}).

\subsection{Tibet solar neutron telescope}
Here, we provide minimal information necessary to understand events reported in this paper;
detailed information on the Tibet SNT, including detection efficiencies for neutrons and 
$\gamma$ rays, is presented in \citet{TibetSNT}. 

SNT installed at the observatory is part of the international solar-neutron observation network. 
It is composed of nine plastic scintillation counters and proportional counters that are placed 
around them.
A plastic scintillation counter contains plastic scintillator blocks of area and thickness of
1 $\mathrm{m^2}$ and 40 cm, respectively. 
Thus, the total area of the plastic scintillators is 9 $\mathrm{m^{2}}$. 
The counter has a $\phi$12.7 cm photomultiplier at the top of the counter for
collecting light emissions originating from incident particles. 

Incident charged particles deposit their energies in the thick plastic scintillators via ionization loss, 
and hence can be readily observed with SNT. Incident neutrons produce recoil ions by 
scattering protons or carbons in the plastic scintillators, 
while $\gamma$ rays produce electrons via Compton scattering or
pair creation. Through these processes, SNT is able to measure neutrons and $\gamma$ rays, 
although it does not differentiate between them.
In addition, output signals from the photomultiplier are fed to the data acquisition system, 
amplified and discriminated at 4 levels, which correspond to energy deposits of an incident particle of 
$>$40, $>$80, $>$120, and $>$160 MeV.  
For each of the nine plastic scintillation counters, 
individual discriminated logical signal is counted by scalers every second.

Proportional counters complement the plastic scintillation counters.
A proportional counter has a length of 330 cm and radius of 5 cm, and contains 90\% Ar and 10\% $\mathrm{CH_4}$.
Thirty proportional counters are placed above the nine plastic scintillation counters, while 
seventy-two proportional ones shield the 4 sides of the plastic scintillation ones.
Therefore, the surrounding counters can be utilized as an anti-counter to separate photons and neutrons from charged particles.
In fact, using the surrounding proportional counter signals in anti-coincidnece, 
the four discriminated counting rates of the central plastic counters are reduced by a factor of $0.2$$-$$0.25$.
\subsection{Electric-field mill}
To measure electric-field variations, two commercial electric-field mills (BOLTEK EFM-100)
were installed on the premises. 
One is mounted on the ground, while the other 
is located on the roof of a central building;
hereafter, denoted as EFM1 and EFM2,
respectively. The two mills are arranged $\sim$25 m apart with a vertical distance of 3.4 m.
Individual output signals are transmitted to the central building with optical cables, directly
fed to PCs, and recorded every 0.1 s as 
the electric field strength in the range $\pm$40 $\mathrm{kVm^{-1}}$
with a resolution of 20 $\mathrm{Vm^{-1}}$.

The electric field strength measured by EFM2 is always higher, by a factor of $\sim$2, 
than that by EFM1. Such an 
enhancement of an electric field is often caused by distortion of local electric field lines because of
obstructions such as a building.  In fact, EFM2 is installed near a corner of the roof of a building, 
and hence are more largely affected by such distortion than EFM1, which is located on the ground 
with few surrounding obstructions.  
Considering this disparity, we use only EFM1 data in this paper.

\section{Observations}
\subsection{overview}
Examining the data over the 2010 rainy season from May to October, 
we visually found 25 events in which electric fields largely deviate from fair-weather states of $<$$100\,\, \mathrm{V\, m^{-1}}$.
Five of them accompanied prolonged count enhancements.  Three of the five events are clearly observed
by either YBJ NM or SNT, lasting for 10$-$20 minutes. Similar events with duration of 10$-$20 minutes 
have been already reported by other measurements (for example, \cite{fuji_torii_2009, armenia_2010}). 
On the other hand, the remaining two events, detected by both YBJ NM and SNT, last for $>$30 minutes. 
Such a long-lasting emission has never been observed.

With two reasons, we selected one of the two events that are detected by both YBJ NM and SNT. 
One is that the event is a fast observation of the longest-duration emission 
among other long-duration events. The other is that 
the selected event clearly correlate with electric fields (as shown later), while
the other has only a poor correlation with electric fields. Although 
statistical significance of the latter event for YBJ NM and SNT($>$40 MeV) were around two times 
higher than the selected event,  
we will have to collect additional ones in order to well understand the nature of 
such a poorly correlated event.
\subsection{Count histories}
\begin{figure}
\includegraphics[scale=0.6]{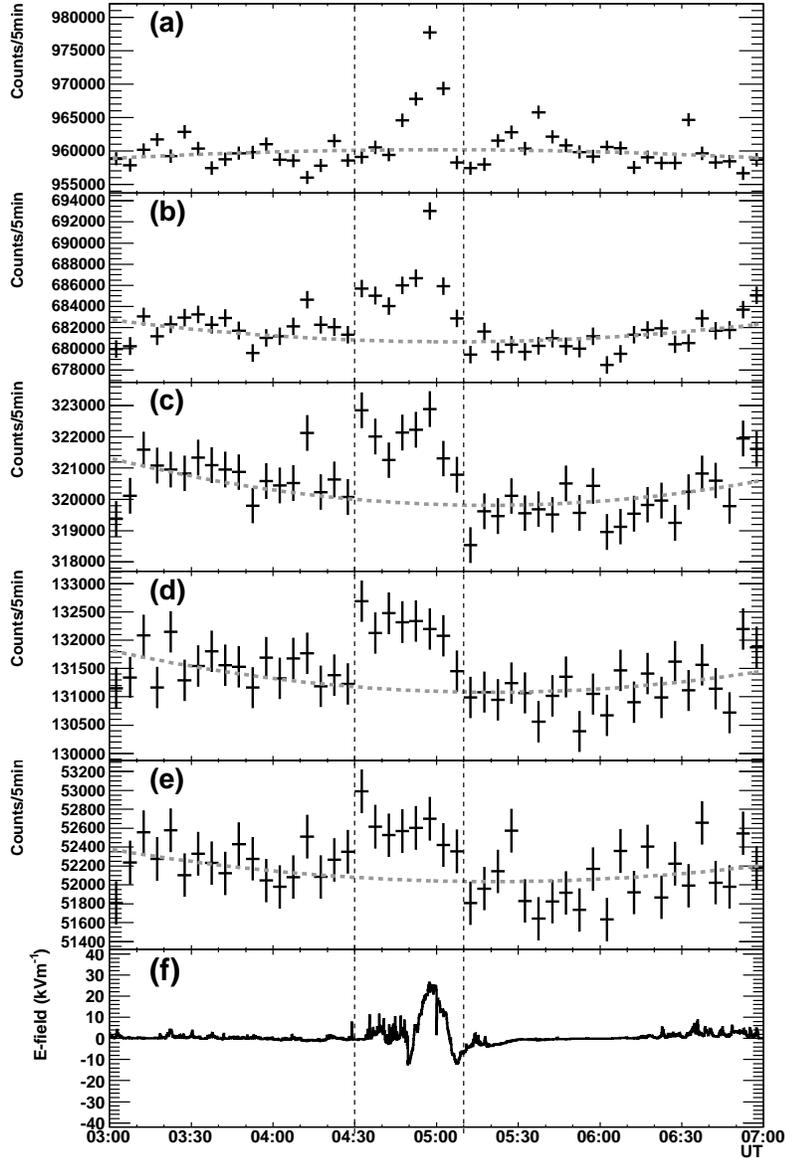}
\caption{%
Count rates per 5 min detected by YBJ NM and SNT, and one-second variations of EFM1 over 3:00$-$7:00 UT on July 22, 2010.
Panel (a) shows count rates by YBJ NM; while panels (b)$-$(e) show count rates by $>$40 MeV, $>$80 MeV, $>$120 MeV, and $>$160 MeV SNT with 
anti-coincidence. Panel (f) shows the one-second variations by EFM1.
Dashed gray curves in panels (a)$-$(e) indicate the estimated background, while vertical dashed lines in all panels represent the defined burst periods.
The horizontal axes show universal time. Error bars are statistical 1$\sigma$ except for panel (f). 
}
\label{fig:20100722_5min}
\end{figure}

Figure~\ref{fig:20100722_5min} shows five-minute counting rates by
YBJ NM and SNT and one-second electric-field variations obtained over 3:00$-$7:00 UT on July 22, 2010. 
All counting rates by YBJ NM and SNT are corrected
for atmospheric pressure variations. In this event, YBJ NM count rates and
SNT ones in $>$40 MeV clearly increase around 5 UT [Figs.~\ref{fig:20100722_5min}(a) and (b)]. 
In addition, higher-energy channels of SNT [panels (b)$-$(e) of Figs.~\ref{fig:20100722_5min}] 
appear to show count enhancements in coincidence with the above clear increases. 
Given the clear signals, in particular, the $>$40 MeV channel of SNT vetoing charged particles 
with the anti-counter [Fig.~\ref{fig:20100722_5min}(b)],  
we can conclude that $>$40 MeV $\gamma$ rays and/or neutrons reach the detectors to produce the observed signals.

With a criterion that individual counts of the $>$40 MeV channel of SNT continuously have $>$$2\sigma$ statistical significance
above background, we define burst time as 40 min at 4:30$-$5:10 UT.
Here, by excluding the data in this period and fitting the remaining data with a quadratic function, 
we estimate the background for YBJ NM and SNT (gray dashed curves in Figs.~\ref{fig:20100722_5min}). 
Subtracting the interpolated background from total observed counts in the burst period, 
we obtain net count increases for the burst recorded by YBJ NM and SNT; these are listed in Table~\ref{tab:increase_two_events}
together with their statistical significance. Hereafter, the burst is simply called 100722. 

Generally, the counting rate of a neutron monitor, including YBJ NM,  does not simply obey Poisson 
distribution because of the multiplication of one incident neutron in the lead blocks. 
\citet{NMnetwork} provide a more detailed explanation on how these effects cause non-Poissonin fluctuations in NM data.
They described that a statistical significance obtained by a NM
usually should be reduced by a factor of 1.2$-$2, depending on the geomagnetic cut-off rigidity
and the atmospheric depth at NM locations. Thus, the statistical significance obtained (Table~\ref{tab:increase_two_events})
may decrease by half. Importantly, both YBJ NM and SNT simultaneously recorded large 
enhancements in association with electric-field variations.

Based on the following features of the event observed, we may conclude that it
is associated with thunderclouds, but not lightning. First is its long duration;
apparently, such a long-duration emission would not be generated by lightning and/or
its related phenomena that generally last for milliseconds or less. Second is that the
electric field strength in the burst period does not change rapidly (within 1 s), 
but gradually [Figs.~\ref{fig:20100722_5min}(e)].
In addition, although not homogeneous, these features have already been reported 
by many groups \cite[][]{Alex_2002,monju_2002,fuji_torii_2009,norikura_tsuchiya_2009,growth_2007,armenia_2010,monjyu_torii_2011,growth_2011}
as thundercloud-related emissions. 

\subsection{Relation with electric fields}
\begin{figure}
\includegraphics[width=0.7\textwidth]{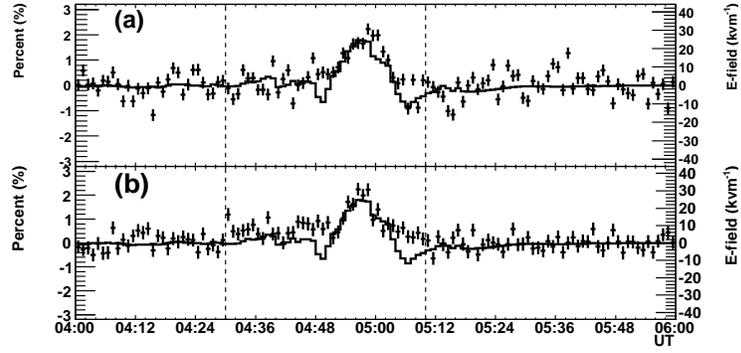}
\caption{%
One-minute count histories in percent observed by YBJ NM and SNT, and one-minute averaged electric-field variations.
Data points with 1$\sigma$ error bars in panels (a) and (b) correspond to the variations in YBJ NM and $>$40 MeV SNT with anti-coincidence 
for 100722, respectively.
In all panels, histograms (solid lines) represent the average field variations
by EFM1. The horizontal axes show universal time. Left and right vertical axes  
denote the count variations in percent and the electric field strength in $\mathrm{kV\,m^{-1}}$, respectively.
Vertical lines in each panel indicate the burst onset and end times.
}
\label{fig:NM_NT_aveEFM_relation}
\end{figure}

\begin{figure}
\includegraphics[width=0.60\textwidth]{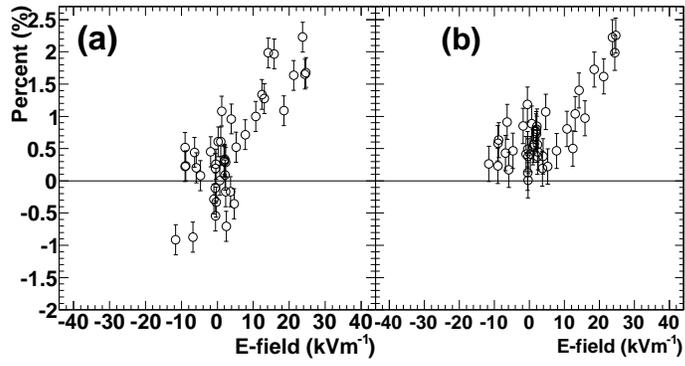}
\caption{%
Burst count variations plotted against electric fields measured by EFM1. Open circles are 
count variations in the burst intervals shown in Fig.~\ref{fig:NM_NT_aveEFM_relation}.
Panels (a) and (b) correspond to YBJ NM and SNT ($>$40 MeV) for 100722, respectively.
}
\label{fig:E_vs_per}
\end{figure}

Figure~\ref{fig:NM_NT_aveEFM_relation} represents detailed time variations of
YBJ NM and SNT,  together with the averaged electric fields. Clearly, peaks of YBJ NM and SNT signals
for 100722, obtained over 4:50$-$5:04 UT, correlate with those of
electric fields in the same interval [Figs.~\ref{fig:NM_NT_aveEFM_relation}(a) and (b)]. 
For clarity, Figure~\ref{fig:E_vs_per} shows the correlation between the present burst and the electric field
measured by EFM1.  We computed a correlation coefficient between the count variations of 
YBJ NM and SNT and the electric field as 0.79 (0.01)  and 0.77 (0.03), respectively. 
Each number in parentheses represents a correlation coefficient outside the burst period.

An electric field in the downward direction is measured as a positive field. Thus, 
the positive electric fields correspond to the existence of positive charges overhead,
which are frequently observed at Tibet~\cite{Qie_2005} and New Mexico~\cite{SM_2008} when 
thunderclouds exist at a mature stage over a field mill on the ground. Furthermore, such a thundercloud
generally forms tripole electrical structures, which consist of positive, negative, and positive layers 
from top to bottom, which in turn accelerates electrons therein toward the ground.
\section{Neutron production and propagation in the air}
\subsection{outline}\label{subsec:outline}
According to ~\citet{Babich_Npro_2010}, a yield rate of a photonuclear neutron per one gamma ray 
with energies $>$10 MeV is $4.3\times 10^{-3}$. Produced neutrons propagate in the atmosphere, 
attenuated by elastic and/or inelastic collisions with air nulcei. Assuming neutrons propagate
over $L=$1 km (0.1 km) to reach the observatory,  neutrons produced decrease in number by a factor
of $\exp{(-L/\lambda_\mathrm{n})}=2\times10^{-3}\,  (0.5)$. Here, $\lambda_\mathrm{n}$ represents an attenuation length of neutrons
in the atmosphere, calculated as $\lambda_\mathrm{n}=13\,\, \mathrm{g\, cm^{-2}}$ for 20-MeV neutrons using 
total cross section between a neutron and an air nucleus~\cite{Shibata_SN_propagation_1994}.
As a result,  a $>$10-MeV $\gamma$ ray is found to produce $10^{-5}$$-$$10^{-3}$ neutrons
to arrive at the observatory. Given this arrival rate of neutrons and derived detections efficiencies for neutrons and
$\gamma$ rays (Fig.~\ref{fig:NM_eff}), we expect that $>$10-MeV $\gamma$ rays would be able to considerably 
contribute to the signals detected by YBJ NM.  To better understand how much photonuclear
neutrons propagate to the observatory, 
we performed a GEANT4 simulation. 

For the purpose of simulating neutron production via the photonuclear reaction and neutron propagation
in the atmosphere, 
we constructed five atmospheric layers starting from the observatory level (4.3 km a.s.l.) to 5 km higher.
Each rectangular atmospheric layer has a vertical length (z direction) of 1 km and horizontal length (xy directions) of 10 km, and consists
of $\mathrm{N}_{2}$, $\mathrm{O}_2$, and $\mathrm{Ar}$ with mole ratios of 
78.1\%, 21.0\%, and 0.9\%, respectively.  Air density in the individual layers is fixed at 
$7.7\times10^{-4}\,\, \mathrm{g\, cm^{-3}}$ for 4.3$-$5.3 km a.s.l., 
$7.0\times10^{-4}\,\, \mathrm{g\, cm^{-3}}$ for 5.3$-$6.3 km a.s.l., 
$6.6 \times 10^{-4}\,\, \mathrm{g\, cm^{-3}}$ for 6.3$-$7.3 km a.s.l.,
$5.6 \times 10^{-4}\,\, \mathrm{g\, cm^{-3}}$  for 7.3$-$8.3 km a.s.l., and
$5.0 \times 10^{-4}\,\, \mathrm{g\, cm^{-3}}$ for 8.3$-$9.3 km a.s.l~\cite{atmos_def}.
In the following simulations, seven source heights are assumed to be 0.3, 0.6, 0.9, 1.5, 2, 3, 
and 5 km above the observatory level. From each source height, one million $\gamma$ rays
were injected to the atmosphere to produce secondary particles. 
The secondary particles, propagating to the observatory, were saved with their
species, energy, x-y positions, azimuth, and zenith angles. 

Bremsstrahlung $\gamma$ rays derived from runaway electrons 
have been thought to have an exponentially cut-off power-law spectrum, with a cut-off energy 
of $\sim$7 MeV~\cite{Dwyer_Eave_2004,Babich_Eave_2004}. However, the recent AGILE observation~\cite{AGILE_TGF100MeV}
indicated that a high-energy part ($>$1 MeV) of the TGF spectrum extending from 10 MeV to 100 MeV 
can be explained by a power-law spectrum with a spectrum index, $\beta$, of $-2.7$ rather than an exponentially cut-off one.
Sea-level observations of long-duration $\gamma$ rays also showed that
a source $\gamma$-ray spectrum may be described as a power-law type
with $\beta \sim -2$~\cite{growth_2011}. Theoretically, $\beta$ of a bremsstrahlung $\gamma$-ray spectrum has the hardest limit of  $-1$.
We therefore assumed a power-law spectrum as an initial photon spectrum in this study 
and $\beta$ is $-1$, $-2$, or $-3$. 
The minimum and maximum energies of the spectrum are set at 10 and 300 MeV, respectively,
to fully cover the presently relevant energy range.
In addition, downward directions of initial $\gamma$ rays were assumed to be distributed either isotropically within 0$-$30 degrees 
or over a Gaussian beam with a half-opening angle of 30 deg. Both types would be expected from runaway electrons
moving in electric fields in air, because moving electrons are subjected to multiple scatterings with air molecules, 
and the geometrical or electrical structure of electric fields in thunderclouds may not be very simple~\cite{RU_lightningbook}.

\subsection{Energy spectrum}\label{subsec:NP}
Figures~\ref{fig:n_spectra_spe_300_iso} and \ref{fig:n_spectra_spe_300_beam} show neutron energy spectra
obtained by the isotropic and Gaussian angular distributions, respectively. 
There is no significant difference in shape of the neutron spectra between the two angular distributions.
These neutron spectra suggest that the neutrons arriving at the observatory have a mean energy of 1$-$10 MeV
and the maximum energy of produced neutrons is about one-third of that
of the $\gamma$ rays emitted from a source.  
The former feature has been reported by \citet{Carlson_Npro_2010} and \citet{Babich_Npro_2010} as well. 

Figure~\ref{appfig:gammarays_electrons_spe} represents spectra for $\gamma$-ray and electrons assuming the isotropic emission of initial $\gamma$ rays.
Similar to neutron spectra, those for $\gamma$ rays and electrons, 
assuming the Gaussian beam emission, do not largely change from the isotropic ones. 

\subsection{Survival rate}
Figure~\ref{fig:n_rate_spe} shows survival rates for $>$1 keV neutrons and $>$10 MeV $\gamma$ rays for the two
angular distributions, sampling arriving neutrons ($\gamma$ rays) with energies of $>$1 keV ($>$10 MeV) and calculating 
a ratio of the number of the arriving neutrons ($\gamma$ rays) to that of primary $\gamma$ rays. 
The threshold energy of 1 keV for neutrons does not affect our results, because 
neutrons with energies $<$1 keV constitute a maximum 5\% of all neutrons produced.
As expected, the neutron survival rates for the two angular distributions are similar in shape and intensity,
having at most 10\% difference in rate. 
Depending on spectrum indices, the neutron survival rates are generally 
constant at $\sim$$10^{-3}$ until the source height is around 1 km, and then decrease to $10^{-4}$$-$$10^{-5}$. 
The derived survival rates quite agree with those simply calculated in Sec.~\ref{subsec:outline}. 

As can be easily seen, each neutron survival rate has its peak at the source height of $\sim$0.6 km, which corresponds to 
$\sim$50 $\mathrm{g\, cm^{-2}}$. 
The shape of the survival rates of neutrons simply reflects the product of
the probability that the photonuclear reaction occurs at the point $\gamma$ rays
propagate in the atmosphere and that the produced neutrons are attenuated, which is
proportional to $[1-\exp{(-H/\lambda_\mathrm{p})}]\times\exp{(-H/\lambda_\mathrm{n})}$.
Here, $H$ represents the assumed source height, while $\lambda_\mathrm{p}$ and $\lambda_\mathrm{n}$
represent the interaction length of $\gamma$ rays to cause photonuclear reaction, 
which is $\sim$3000 $\mathrm{g\, cm^{-2}}$ at the peak cross section of 15 mb, and the attenuation length of neutrons, respectively.
$\lambda_\mathrm{n}$ in the relevant neutron energies of 1$-$100 MeV 
range between 20$-$100 $\mathrm{g\, cm^{-2}}$, corresponding to 0.2$-$1.4 km.

\section{Contribution ratios to the signals}\label{sec:dis_cntribution}
\subsection{Method}
\begin{figure}
\includegraphics[width=0.4\textwidth]{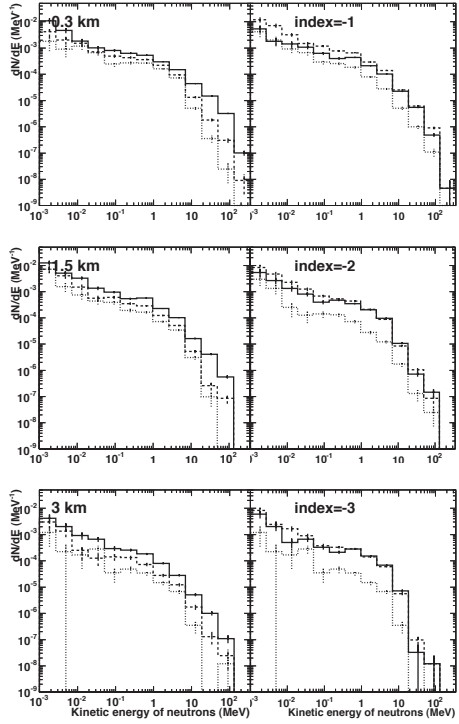}
\caption{%
Derived energy spectra of neutrons reaching the observatory level, assuming the isotropic emission
of initial $\gamma$ rays.
Left panels show those spectra with an initial source height fixed at a given value in each panel.
Solid, dashed, and dotted lines in the left panels correspond to $\beta$ of $-1$, $-2$, and $-3$, respectively. 
Right panels indicate those spectra in which $\beta$ is fixed at a constant value with source heights of 0.3 (solid), 
1.5 (dashed), and 3 km (dotted). Each horizontal axis denotes
kinetic energy of neutrons in MeV at the observatory. The vertical axes indicate relative values divided by the number of incident
$\gamma$ rays $1\times10^6$.
}
\label{fig:n_spectra_spe_300_iso}
\end{figure}

\begin{figure}
\includegraphics[width=0.4\textwidth]{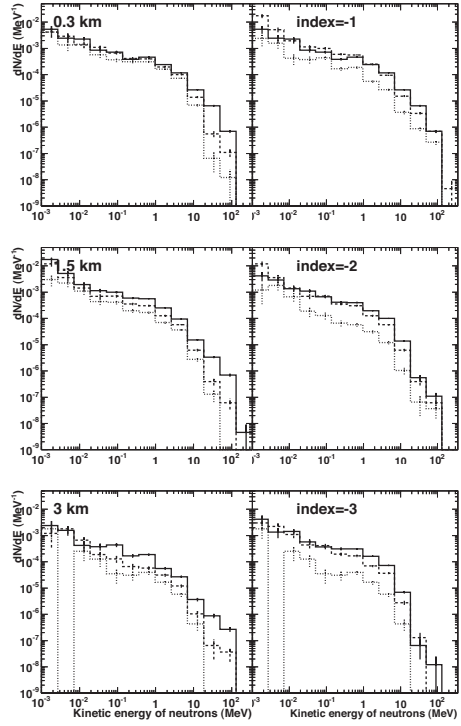}
\caption{%
Same as Fig.~\ref{fig:n_spectra_spe_300_iso}, but for Gaussian-type angular distribution of initial $\gamma$ rays.}
\label{fig:n_spectra_spe_300_beam}
\end{figure}

\begin{figure}
\includegraphics[width=0.4\textwidth]{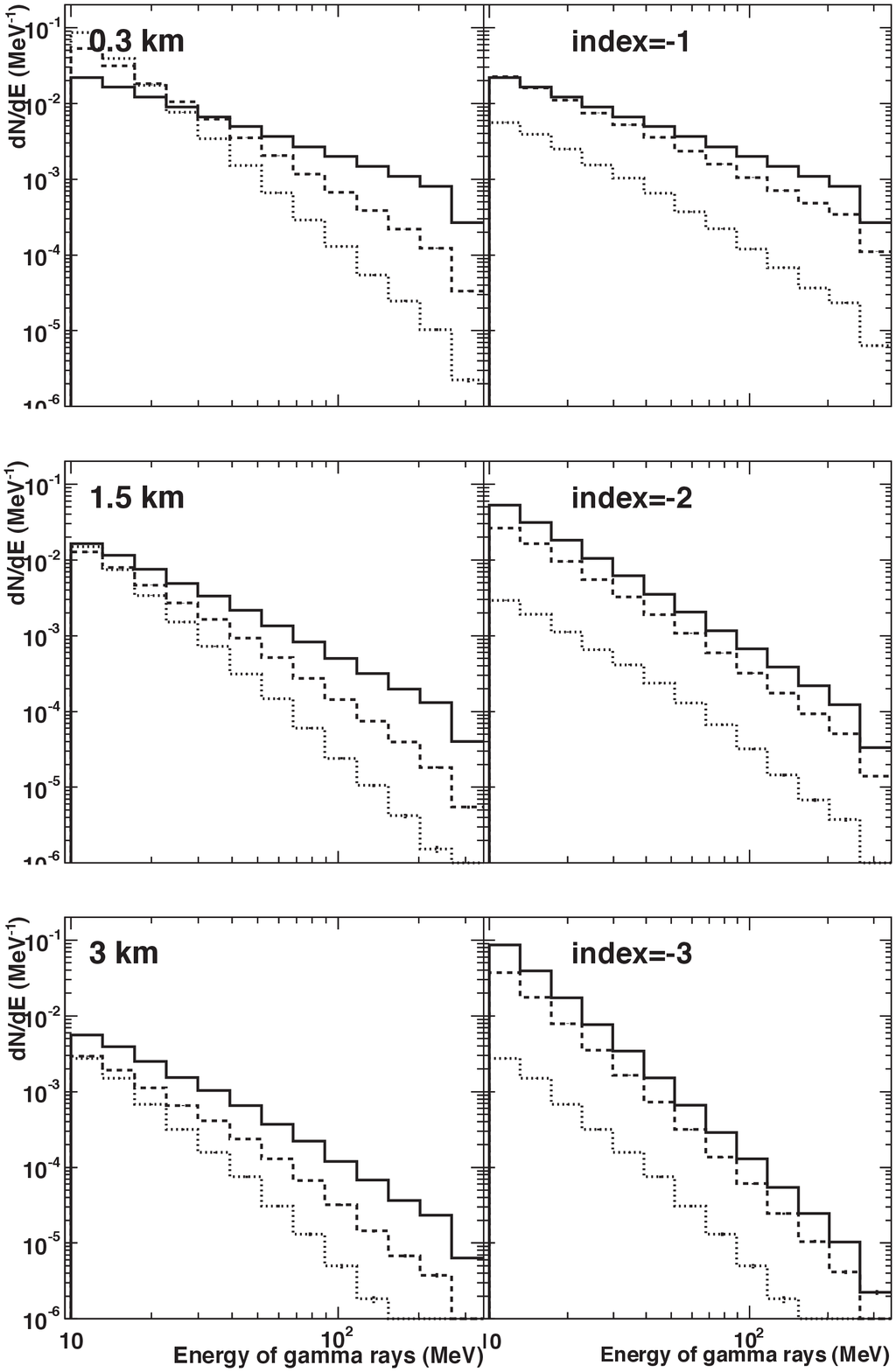}
\includegraphics[width=0.4\textwidth]{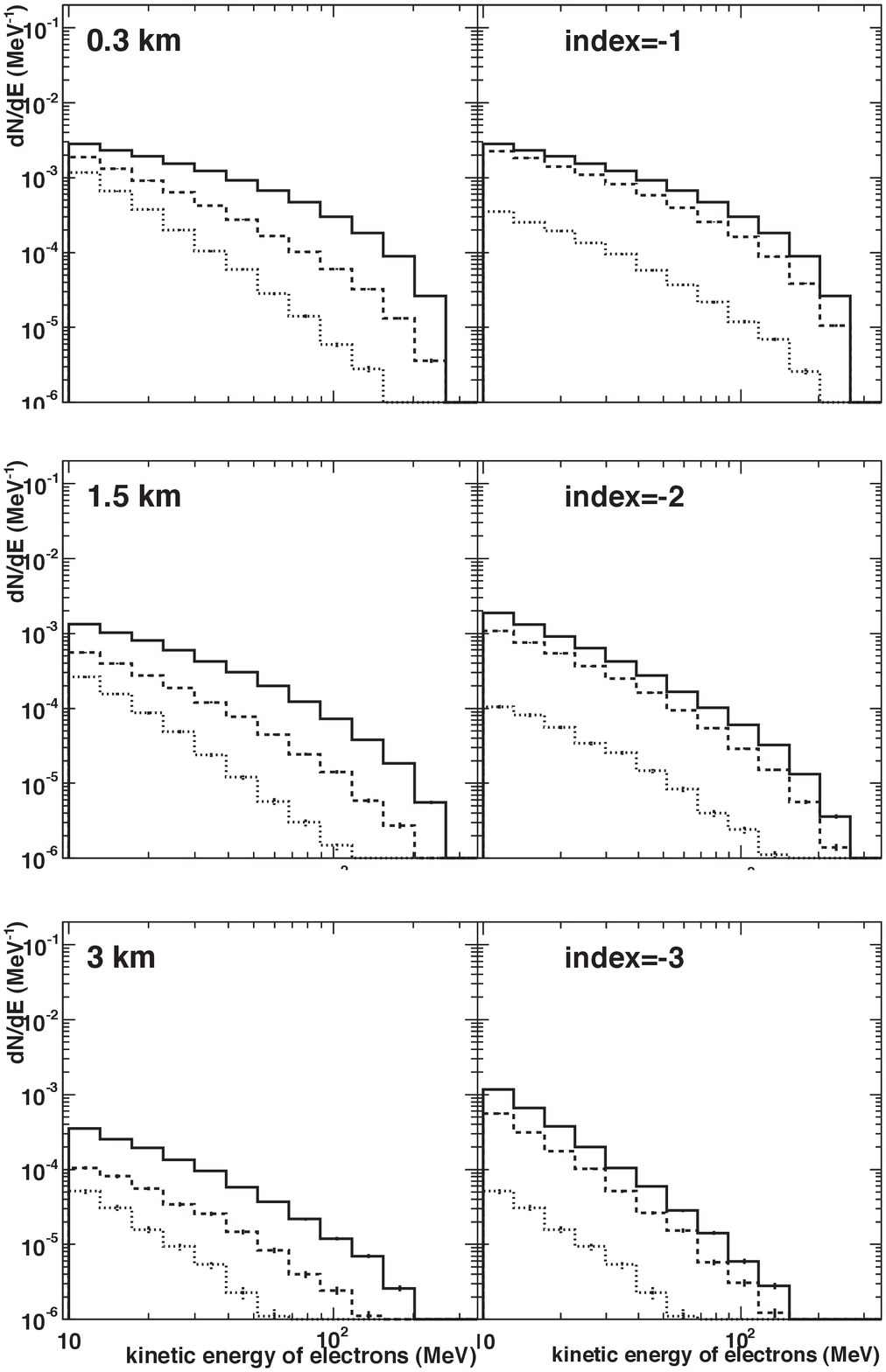}
\caption{Same as Fig.~\ref{fig:n_spectra_spe_300_iso}, but for gamma rays (left) and electrons (right).}
\label{appfig:gammarays_electrons_spe}
\end{figure}

Given the simulated neutron spectra (Figs.~\ref{fig:n_spectra_spe_300_iso} and \ref{fig:n_spectra_spe_300_beam}), 
and those of $\gamma$ rays and electrons (Fig.~\ref{appfig:gammarays_electrons_spe}), 
as well as the detection efficiencies of the neutron monitor (Fig.~\ref{fig:NM_eff}), we can examine how neutrons and electromagnetic components
contribute to signals that are expected to be detected by YBJ NM and SNT during thunderstorms.

As argued so far, we presume that the four components, neutrons, $\gamma$ rays, electrons, and positrons,
explain the count increases observed by YBJ NM, and that neutrons and $\gamma$ rays contribute
to SNT signals because SNT utilizes the anti-counter to reject charged particles. 
Therefore, a predicted count increase at a given time $t$ for individual particles, $\Delta n_i (t)$, is written by  
\begin{equation}
\Delta n_i (t) = \alpha A \int_{K_1}^{K_2} I(t) S_i (E_i)\epsilon_i(E_i)dE_i ,
\label{eq:delta_n}
\end{equation}
assuming that the relevant particles have the same production history, $I(t)$, to be generated during thunderstorms. 
Here, $\alpha$ represents a normalization factor with the unit $\mathrm{MeV^{-1}s^{-1}m^{-2}}$ for a source spectrum and $A$ represents 
the area of  YBJ NM (32 $\mathrm{m^2}$) or SNT (9 $\mathrm{m^2}$). $E_i$ denotes the energy of a particle type $i$, $S_i(E_i)$ represents
the spectra (Fig.~\ref{fig:n_spectra_spe_300_iso} for isotropic emissions), and $\epsilon_i(E_i)$ denotes
the detection efficiencies of YBJ NM (Fig.~\ref{fig:NM_eff}) or SNT (Fig. 5 of ~\cite{TibetSNT}). In the present study, $K_1$ 
is set to 1 keV for neutrons and 10 MeV for electromagnetic components, while $K_2$ is fixed at 300 MeV for all components.
By integrating Eq.(\ref{eq:delta_n}) over a certain time interval of $t_2$$-$$t_1$, we can obtain an expected net count increase
due to each particle ($\Delta N_i $) as
\begin{equation}
\Delta N_i = \int_{t_1}^{t_2} \Delta n_i(t)dt.
\label{eq:expcnt}
\end{equation}

Under the present assumption, the simulated spectra $S_\mathrm{i}(E_\mathrm{i})$ in
Eq. (\ref{eq:delta_n}) are independent of time $t$. 
Accordingly, a ratio of $\Delta N_i/\sum \Delta N_i$ is calculated as 
\[ \frac{\Delta N_i}{\sum \Delta N_i}=\frac{\int_{K_1}^{K_2} S_i (E_i)\epsilon_i(E_i)dE_i }{ \sum \int_{K_1}^{K_2} S_i (E_i)\epsilon_i(E_i)dE_i }, \]
which shows the contribution fraction of each species to an expected signal. 
\subsection{YBJ NM signals}
Figure~\ref{fig:NM_cntratio} depicts contribution ratios of neutrons and $\gamma$ rays for YBJ NM,
assuming an isotropic angular distribution. As expected, contribution ratios for the Gaussian angular distribution
are almost the same as those for the isotropic distribution. 
For clarity, contribution ratios for electrons and positrons are not shown. 
Interestingly, the contribution ratios of neutrons and $\gamma$ rays do not depend largely on $\beta$.
Therefore, it is obvious that $\gamma$ rays dominate (96\% to 85\%) the fraction of the expected count increase
as the source is farther, while neutrons contribute a maximum of 15\%.  
\subsection{SNT signals}
Similarly, contribution ratios for SNT signals can be calculated using detection efficiencies for neutrons and $\gamma$ rays
in Fig. 5 of \citet{TibetSNT}.  The Ninety-nine percent of the observed signal for $>$40 MeV channel of SNT is dominated by $\gamma$ rays, 
while the remaining three higher energy channels are almost fully contributed by $\gamma$ rays. These results for SNT 
are mainly ascribed to a relatively small fraction ($<$5\%) of neutrons produced in $>$40 MeV energies
via the photonuclear reaction (Figs.~\ref{fig:n_spectra_spe_300_iso} and \ref{fig:n_spectra_spe_300_beam}).

\section{Time histories of YBJ NM and SNT}\label{subsec:TP}
$I(t)$ can be naturally assumed to follow the time history of the electric field 
[Fig.~\ref{fig:20100722_5min}(f)].
Utilizing the one-sec electric-field variations as $I(t)$, Eq.(\ref{eq:delta_n}) has only one unknown parameter, $\alpha$,
that needs to be determined by comparing an expected time profile of YBJ NM or SNT to the observed one-minute 
profile. Further, we test the following two hypotheses. First, the relevant particles 
are produced only when the electric field at the surface has positive polarity and
electrons in thunderclouds are accelerated toward the ground. Second, 
the particles are generated when the field has negative and positive polarities.

For the purpose of introducing the mathematical form of $I(t)$ for the first assumption, 
the positive electric field strength [Fig.~\ref{fig:20100722_5min}(f)]
in the burst periods is divided by the maximum strength of $26.8 \,\mathrm{kVm^{-1}}$, 
and the negative electric field strength and that outside the individual burst periods are set to zero.  
On the other hand, absolute values of the electric field strength 
divided by the above mentioned maxima, are considered as $I(t)$ for the second assumption. In addition, $I(t)$ in this case 
is zero outside the burst times. Therefore, $\int I(t) dt$ for both assumptions is normalized to one. Hereafter, we call the first and the second
assumptions "negative emission" and "bipolar emission", respectively. 
Substituting each $I(t)$ in Eq. (\ref{eq:delta_n}) and integrating Eq. (\ref{eq:expcnt}) every 60 s
in individual burst intervals, we can prepare a one-minute expected time profile depending on each $I(t)$ and 
compare it with the observed one-minute counting rate of YBJ NM and SNT (Fig.~\ref{fig:NM_NT_aveEFM_relation}). Next, we 
compute
\[
\chi^2 = \sum_i \left [ \frac{N_\mathrm{ob}(t_i) - N_\mathrm{ex}(t_i)}{\sigma_\mathrm{ob} (t_i)} \right ]^2
\]
to search for the $\chi^2$ minimum with $\alpha$ being a free parameter. Here, $N_\mathrm{ob}(t_i)$ and $N_\mathrm{ex}(t_i)$
represent the observed one-minute counts of YBJ NM or SNT and the model-predicted counts at a given time $t_i$ in
the burst intervals, respectively. Statistical errors associated with $N_\mathrm{ob}(t_i)$ are written 
as $\sigma_\mathrm{ob} (t_i)$. Summation was carried out over each burst interval.

In fact, each $I(t)$ produces the same shape of predicted count history and the same $\chi^2$ minimum
for YBJ NM or SNT, despite using simulated spectra [$S_\mathrm{i}$ in Eq.(\ref{eq:delta_n})] obtained
with various sets of $\beta$ and $H$. 
This is because Eq. (\ref{eq:delta_n}) has only one unknown parameter $\alpha$, and the shape of 
$S_\mathrm{i}$ in Eq. (\ref{eq:delta_n}) is independent of $t$. 
To concretely determine $\alpha$, we first independently evaluated $\alpha$ for YBJ NM and SNT with each $I(t)$ by the above method, 
using simulated spectra obtained by 21 combinations of ($\beta$, $H$); $\beta$ of $-1$, $-2$, $-3$ and $H$ of 0.3, 0.6, 0.9, 1.5, 2, 3, and 5 km.
The derived $\alpha$ for YBJ NM and SNT are shown in Figure~\ref{fig:NM_NT_compalpha}. 
Next, subtracting $\alpha$ acquired from YBJ NM data with a set of ($\beta$, $H$) from that acquired from SNT data 
with the same set of ($\beta$, $H$), we searched for the smallest difference in $\alpha$ obtained by the two independent detectors. 

As clearly seen in Figure~\ref{fig:NM_NT_compalpha}, a difference in $\alpha$ for YBJ NM and SNT
is the smallest at ($\beta$, $H$)=($-2$, 900 m) and ($-2$, 600 m) for the negative emission and
bipolar emission, respectively.
Table~\ref{tab:chi2_alpha} displays the calculated $\alpha$ and $\chi^2$ minima, together with a set of $\beta$ and $H$. 
Figures~\ref{fig:20100722_NMNT_TPcomp} compare the observed count histories of YBJ NM and SNT
with those expected from the parameters listed in Table~\ref{tab:chi2_alpha} under the two emissions.
The $\chi^2$ values (Table~\ref{tab:chi2_alpha}) clearly suggest that the observed time profiles 
are reproduced by the negative emission rather than the bipolar emission.
\section{Discussion\label{sec:dis}}
\subsection{Gamma-ray emissions}\label{subsec:gamma_emission}
\subsubsection{Characteristics of $\gamma$-ray emissions}
The present study revealed that high-energy $\gamma$ rays with energies $>$40 MeV originate
from summer thunderstorms. According to sea-level observations in winter thunderstorms~\cite{growth_2011}, 
long-duration $\gamma$ ray emissions from winter thunderstorms extend to 10$-$20 MeV. This may be due to a difference in atmospheric 
density at ground and high mountains. In fact,  a TGF spectrum averagely extends to few tens of MeV~\cite{AGILE_2010,FERMI_2010}, or 
100 MeV on rare occasions~\cite{AGILE_TGF100MeV}.  It is believed that TGFs occur at altitudes of 15$-$20 km~\cite{DS_MC_2005}.
These results including the present one may imply that a lower atmospheric density is attributable to a higher $\gamma$-ray emission.

Compared with other thundercloud-related $\gamma$-ray events, 
the duration of the present event is exceptional with its long duration of 40 min.
Wash-out radioactive radon and its decay products frequently cause count increases in ground-based detectors. 
In addition, a duration of such a radon effect is around 20$-$30 min corresponding to their half-lives. 
Thus, the duration of the radon effect is similar to the present one. However, 
the radon families generate $<$3 MeV $\gamma$ rays, being unable 
to give signals in YBJ NM and SNT. 

According to electric field measurement at the Tibet plateau (4.5 km a.s.l.)~\cite{Qie_2005} and 
a mountain in New Mexico (3.2 km a.s.l.)~\cite{SM_2008}, 
the mature stage of summer thunderclouds seems to last for $\sim$1 h. 
In addition, the measurement in New Mexico revealed that a vertical potential relative to the surface in the mature stage is quasi stable
which are required for electrons to be continuously accelerated in thunderclouds
in order to produce prolonged $\gamma$-ray emissions. 
Therefore, we infer that the present event is mainly associated with the mature stage
of the Yangbajing thunderstorms.
On the other hand, mature stages of winter thunderstorms at a costal area
of Japan sea last for $<$10 min~\cite{KM_1994}. 
In fact, all thundercloud-related $\gamma$ rays observed in winter lasted for at most a few minutes~\cite{monju_2002,growth_2007,monjyu_torii_2011,growth_2011}.
Thus, it is deduced that the longevity of the mature stage plays an important role in determining the duration of thundercloud-related $\gamma$ rays.

From the $\gamma$-ray emission of the present event, 
a source height $H$ was estimated as $H=900$ m (Table~\ref{tab:increase_two_events}), 
giving the source altitude of 5.2 km a.s.l. \citet{Qie_2005} reported that a cloud base 
of summer thunderclouds above the Tibetan plateau (4.5 km a.s.l.) is generally located at $\sim$1 km. In addition, 
\citet{Marshall_2005} clearly showed that a bottom positive layer of a summer thundercloud in
New Mexico is located at 4.5$-$5.5 km a.s.l. Thus, the source altitude of 5.2 km a.s.l. is 
in good agreement with altitudes of the cloud base and the positive bottom layer obtained from these observations. 

As clearly seen in Figure~\ref{fig:20100722_5min}, time structures for YBJ NM and SNT are different 
with each other. In particular, YBJ NM showed no count increases at the burst onset, while 
all the SNT channels ($>$40 MeV to $>$160 MeV) provided count enhancements in 5$-$10 minutes
after the onset. These peculiar time structures might be caused by moving of thunderclouds and 
limited illumination of higher-energy part of bremsstrahlung gamma rays emitted from thunderclouds.  
Actually, it is confirmed that long-duration gamma rays move with thunderclouds~\citep{growth_2011}. In addition, 
the bremsstrahlung gamma rays, especially gamma rays with an energy being close to that of accelerated electrons, 
would be relativistically beamed into a narrow cone. For example, a half opening angle of the cone, $\theta \sim 1/\gamma$, is $0.1^{\circ}$
for a 300 MeV electron, where $\gamma$ is Lorentz factor. Given $H=900$ m, we can obtain
a radius of the gamma rays arriving at the observatory as at most 1.6 m ($900$ m$\times \tan{0.1^{\circ}}$).
Because YBJ NM is located $\sim$10 m apart from SNT, $>$40 MeV gamma rays moving with thunderclouds 
might not happen to face towrad YBJ NM over the burst onset.

\subsubsection{Electric potential}
Due to ionization loss of electrons, an electric potential of 40 MV is not high enough to accelerate electrons
to 40 MeV. In practice,  an electric field strength of 240$-$270 $\mathrm{kV\, m^{-1}}$ is
required for electrons of 1$-$10 MeV to 
be accelerated to 40 MeV assuming a vertical length of a high-electric field region is 0.5$-$1 km, as
determined by balloon experiments~\cite{Eack_1996,Marshall_2005}. Multiplying this field strength by
the assumed vertical length, the electric potential of at least 120 MV must be established in the thunderclouds. 
This value of 120 MV is approximately equal to the maximum potential of 130 MV observed by balloon soundings~\cite{MS_Efield}. 
In addition, the AGILE observation of TGFs showed that the electric potential in thunderstorms
is on the order of 100 MV~\cite{AGILE_TGF100MeV} over macroscopic lengths such as cloud sizes or intracloud distances. 
Accordingly, the present observations may show manifestation of the highest potential field during thunderstorms.

\subsubsection{Avalanche multiplication factor}
In addition to quasi-stable electric fields, a stable or quasi-stable source of seed electrons 
would generally be needed for prolonged $\gamma$-ray emissions.
\citet{Gurevich_1992} originally postulated that secondary cosmic rays consist of seed electrons, which increase
in number and emit bremsstrahlung $\gamma$ rays. Thus, according 
to this premise, we derive an avalanche multiplication factor, $M$, expected from the RREA mechanism.

Using $\alpha$ and $\beta$ for the negative emission (Table \ref{tab:chi2_alpha}), 
a source $\gamma$-ray spectrum, $F_\mathrm{s}(E)$, can be described
as $F_\mathrm{s}(E) = \alpha_\mathrm{w} E^{-2}$.
Here, $E$ is a photon energy in MeV and $\alpha_\mathrm{w} = (4.3 \pm0.2)\times 10^3\,\, \mathrm{m^{-2}s^{-1}MeV^{-1}}$ 
is a weighted mean calculated by the values of $\alpha$ from YBJ NM and SNT.
Using $F_\mathrm{s}$ and the burst duration $\Delta T = 2400$ s, we 
estimated the total number of electrons with 10$-$100 MeV energies as 
$N_\mathrm{e} \sim 10^{14}$, in the same manner as estimated by \citet{growth_2011}. For this purpose
we assumed a single acceleration region in the thundercloud with the vertical length and horizontal one of $Z=$ 500 m or 1000 m and
$L=$ 600 m~\cite{growth_2011}, respectively. In reality, a positive or a negative charge layer of thunderclouds may consist of multi 
cells~\cite[e.g.][]{3Dstructure_2010} to form several acceleration area therein. 
Thus, the single acceleration region is a simple assumption to consider individual particle accelerations. 

The secondary cosmic-ray electron flux above 1 MeV at the relevant altitude 
 is $I_\mathrm{0}\sim 400$ $\mathrm{m^{-2}\ s^{-1}}$~\cite{CR_text}.
Therefore, the number of such electrons $N_\mathrm{cr}$ entering the acceleration region in the burst period is computed as
\[
N_\mathrm{cr} = I_\mathrm{0} \times A_\mathrm{s} \times \Delta T \sim 3 \times10^{11} \left( \frac{L}{600\, \mathrm{m}} \right)^2, 
\]
giving $M$ as 
\[
M=N_\mathrm{e}/N_\mathrm{cr}=300\left( \frac{600 \, \mathrm{m}}{L} \right)^2. 
\]
Furthermore, based on the RREA mechanism, $M$ thus derived is described as 
\begin{eqnarray}
M           &=\exp{(Z/\lambda)}, \,\,
\lambda &=\frac{7300\,\, \mathrm{kV}}{V - 276(P/1\,\, \mathrm{atm})}, 
\label{eq:M}
\end{eqnarray}
where $\lambda$ and $V$ denote a length parameter given by \citet{Dwyer_RF_2003}
and electric field strength in $\mathrm{kV \,m^{-1}}$, respectively. Substituting $M= 300$, $Z=500$ or 1000 m, 
and $P = 0.55$ atm (average pressure at $H=900$ m) in Eq. (\ref{eq:M}), we obtain $V = 240$ and 190 $\mathrm{kV\, m^{-1}}$ 
for $Z = 500$ m and 1000 m, respectively. These values of $V$ are consistent with the above estimated
field strength to accelerate electrons to 40 MeV or higher energies. 

Conducting sea-level observations in winter, \citet{growth_2011} showed
that secondary cosmic-ray electrons are multiplied by a factor of 3$-$30 to produce thundercloud-related $\gamma$ rays. 
On the other hand, \citet{armenia_2010} obtained a multiplication factor of $\sim$330 with a
high-mountain measurement in summer. From these results as well as
our result, a multiplication factor in high mountains can be considered to be different from that at sea level.  However, 
the above $M$ becomes 30 if $L=2$ km, which is observed as the horizontal extent of a bottom positive
layer in a summer thundercloud~\cite{Marshall_2005}.
Thus, if $L$ is longer than 2 km, the estimated $M$ may become consistent with that derived from sea-level observations.

\subsection{Neutron emissions}%
\subsubsection{Comparison with the Aragats neutron monitor}
Similar to the present event, \citet{armenia_2010} demonstrated that a count enhancement lasting for 10 min was detected by 
the Aragats neutron monitor located at 3250 m a.s.l. As a result, they concluded that 
the observed increase is fully attributable to neutrons related to the photonuclear reaction.
On the other hand, the present results demonstrate that $>$10-MeV $\gamma$ rays dominate the signals 
observed by YBJ NM. 
This is a main difference between the present study and that by \citet{armenia_2010}.

\begin{figure}
\includegraphics[width=0.4\textwidth]{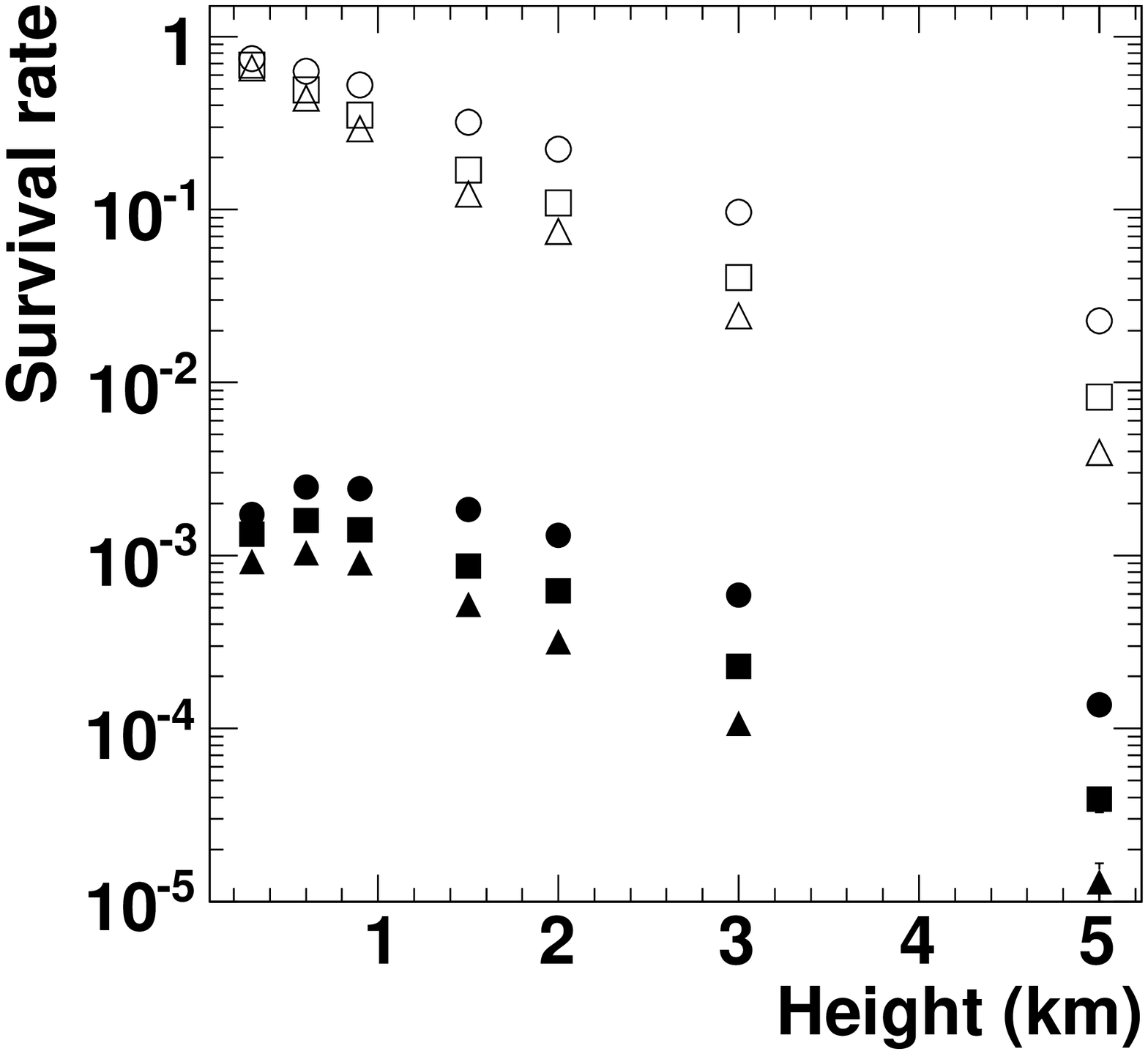}
\includegraphics[width=0.4\textwidth]{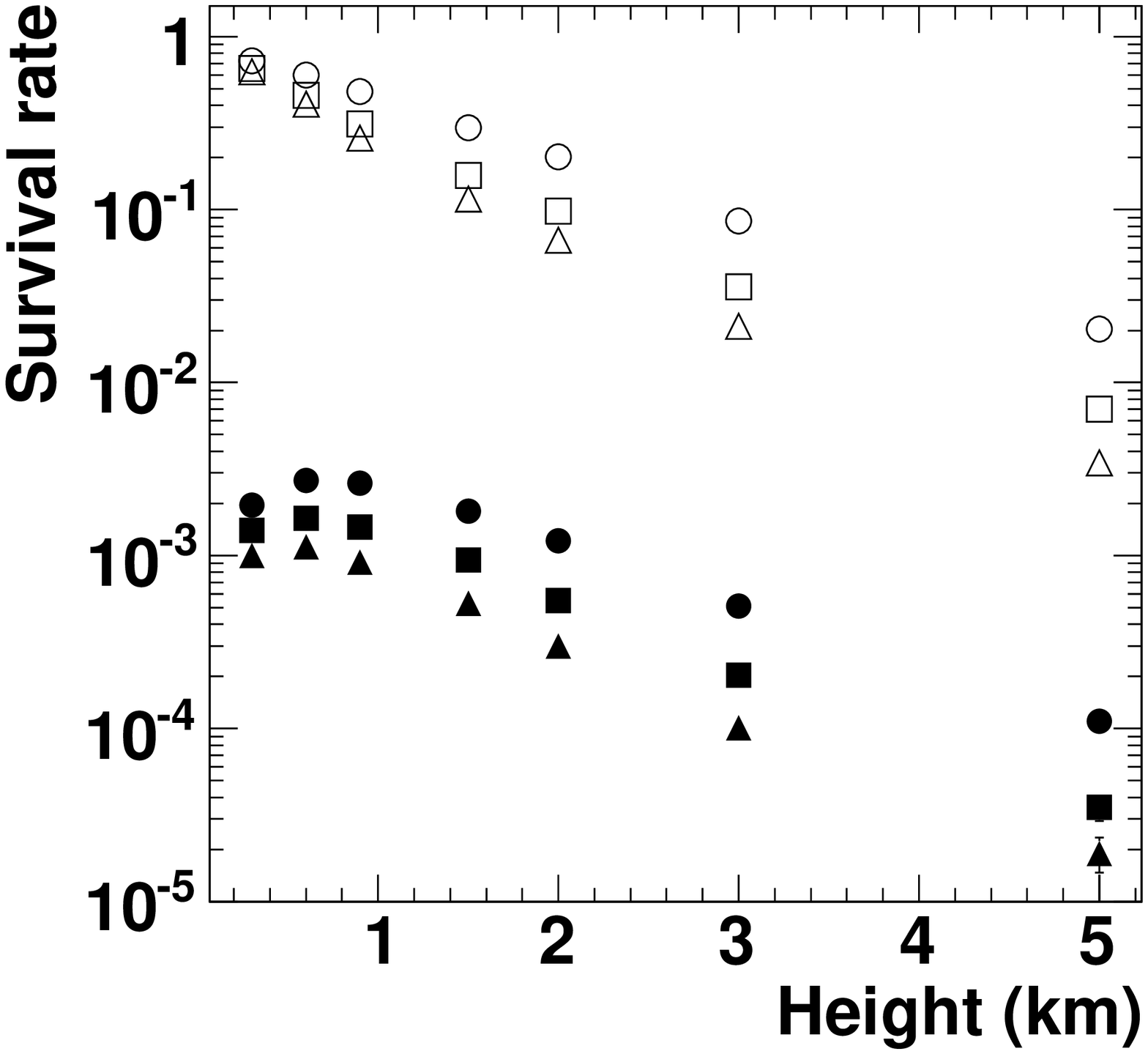}
\caption{%
Survival rates of $>$1-keV neutrons (filled) and $>$10-MeV $\gamma$ rays (open) at
the observatory level, obtained by the $\gamma$-ray power law spectrum simulations.
Left and right panels indicate the isotropic and Gaussian beam distributions of
initial $\gamma$ rays, respectively. Circles, squares, and triangles correspond to $\beta$ of $-1$, $-2$, and $-3$, respectively.
The horizontal axes show source heights.  Errors are statistical 1$\sigma$ errors.
}
\label{fig:n_rate_spe}
\end{figure}

\begin{figure}
\includegraphics[width=0.9\textwidth]{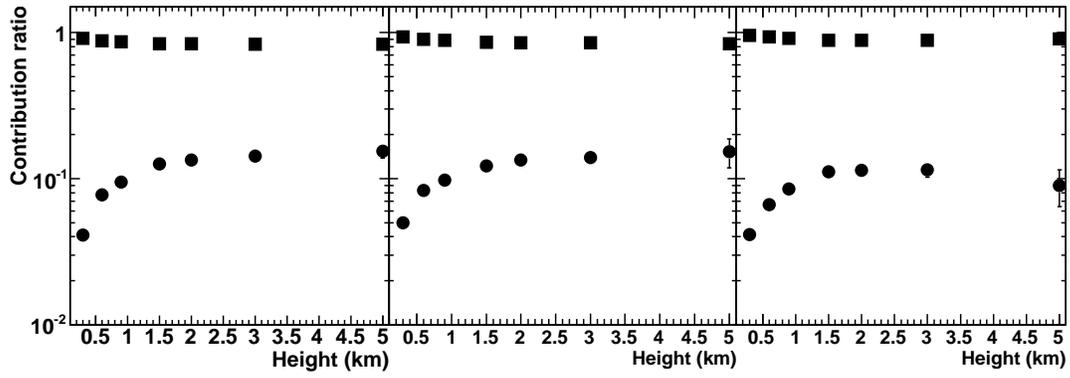}
\caption{%
Contribution ratios of $\gamma$ rays (squares) and neutrons (circles) for YBJ NM signals, 
 plotted against assumed source heights.  Left, middle, and right panels correspond to 
 $\beta$ of $-1$, $-2$, and $-3$, respectively. Errors  are statistical 1$\sigma$ errors.
}
\label{fig:NM_cntratio}
\end{figure}

\begin{figure}
\includegraphics[width=0.4\textwidth]{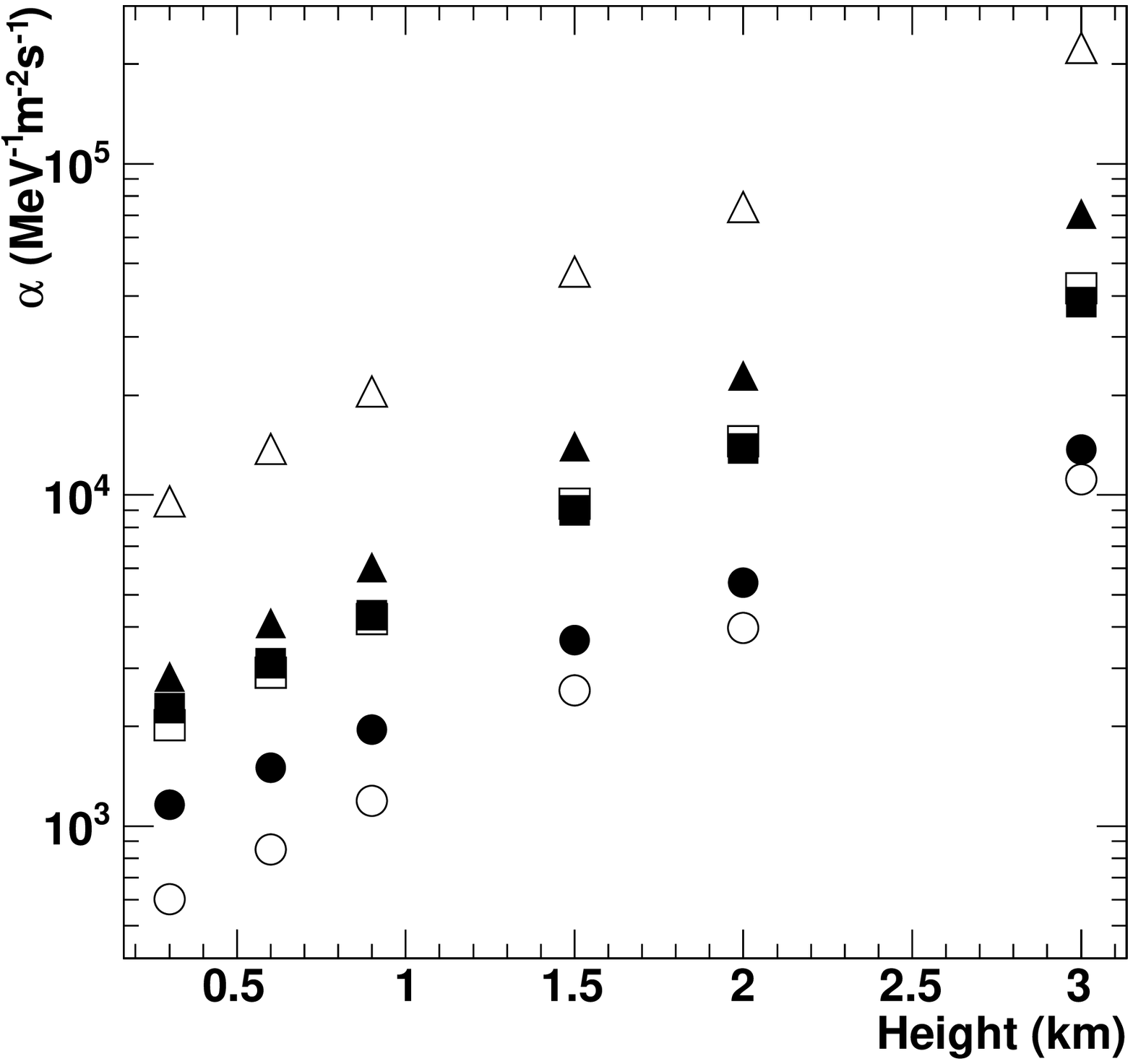}
\includegraphics[width=0.4\textwidth]{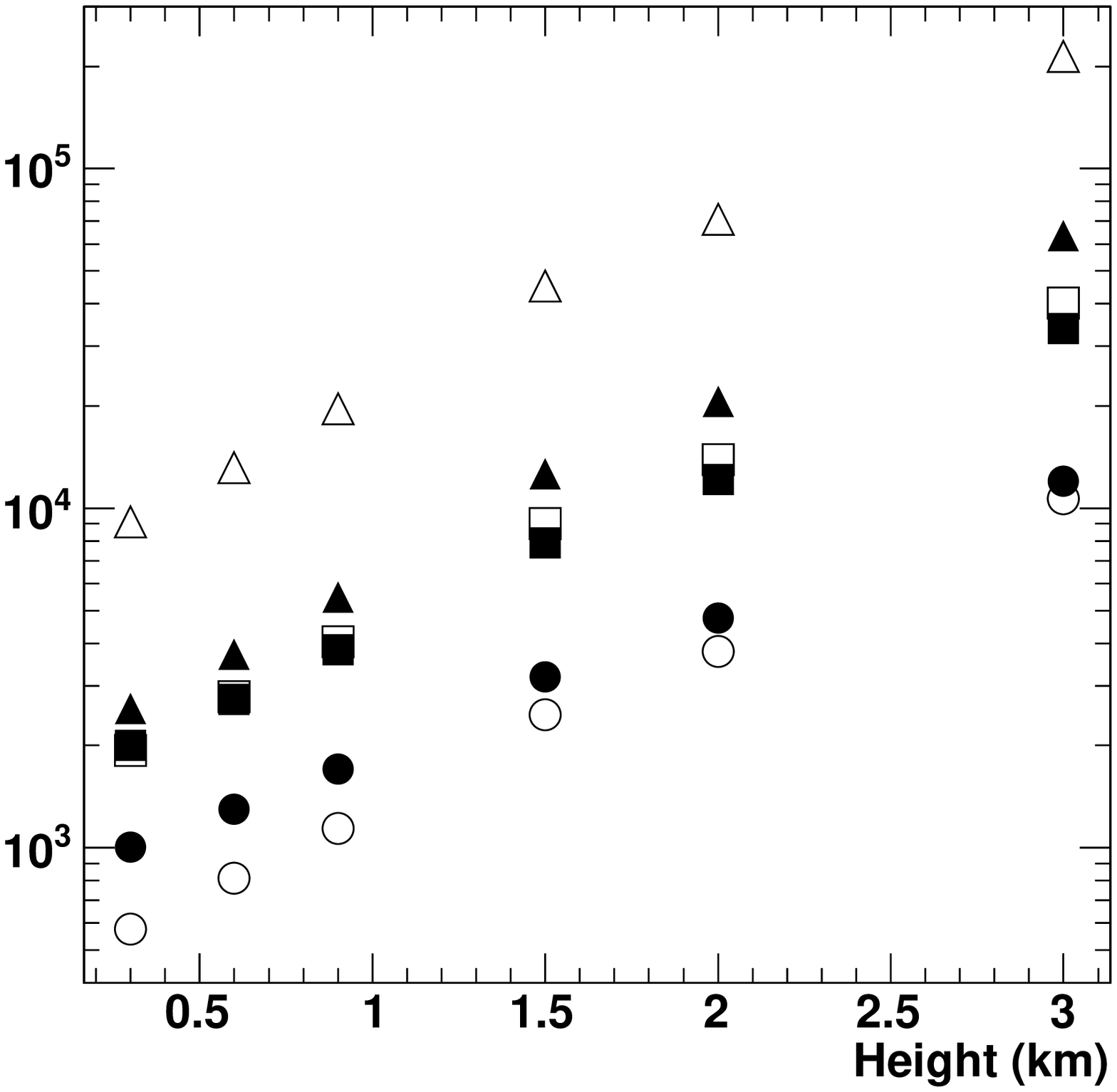}
\caption{%
Comparison of derived $\alpha$ for YBJ NM (open symbols) and SNT (filled symbols). 
Left panel shows the negative emission, while right one denotes the
bipolar emission. Circles, squares, and triangles correspond to $\beta$ of $-1$, $-2$, and $-3$, respectively.
The horizontal axis in each panel represents assumed source height in km.
}
\label{fig:NM_NT_compalpha}
\end{figure}

\begin{figure}
\includegraphics[width=0.45\textwidth]{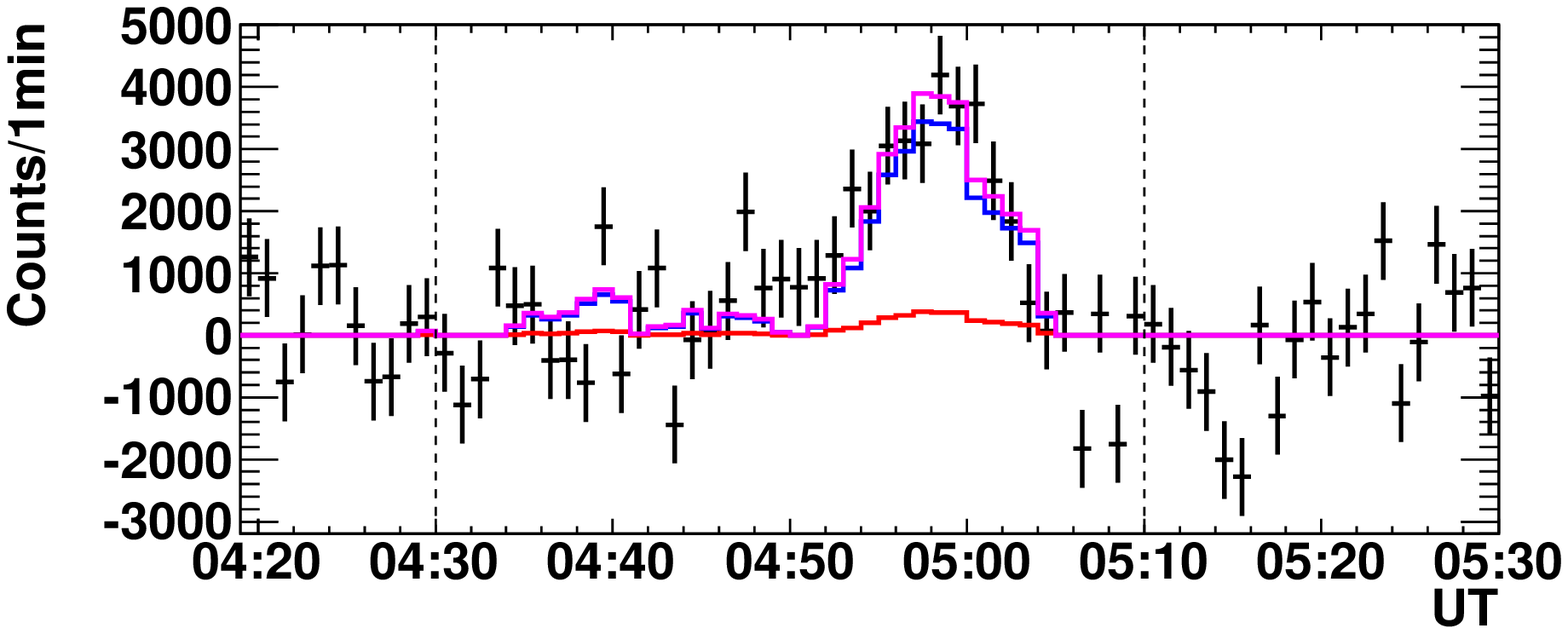}
\includegraphics[width=0.45\textwidth]{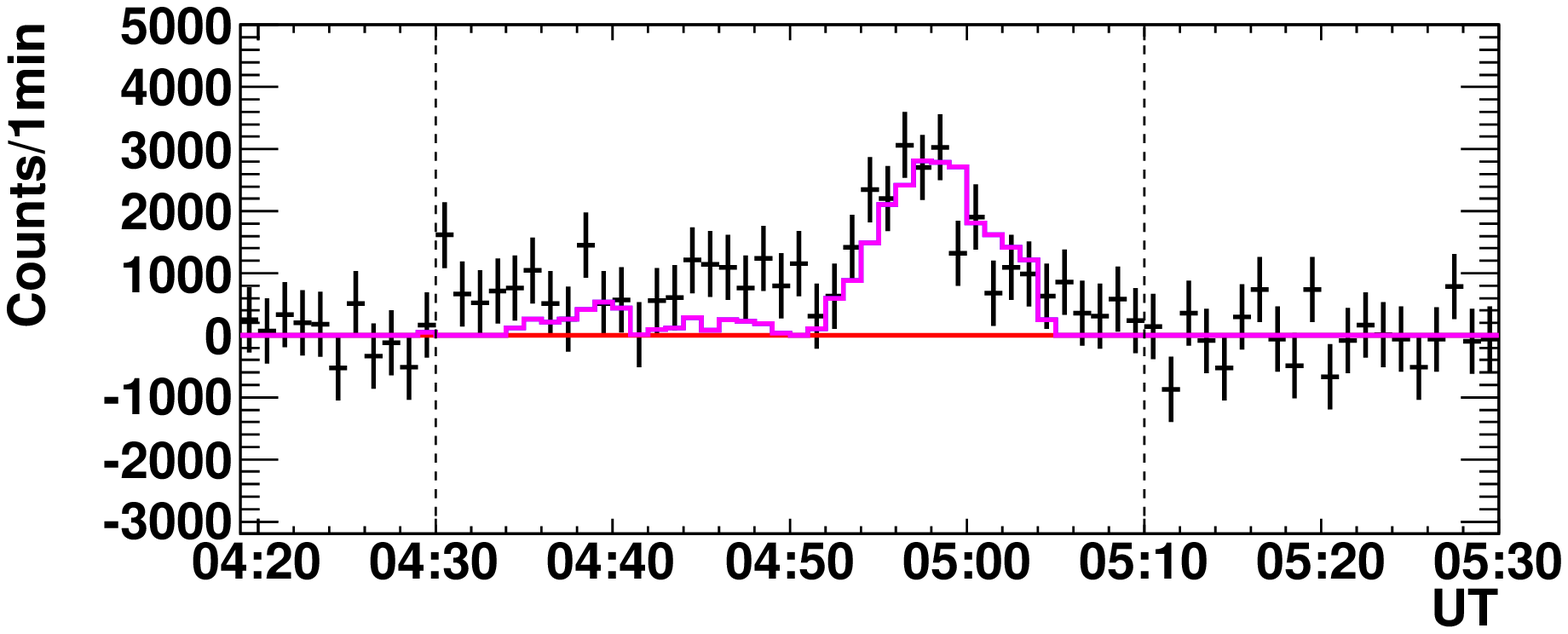}
\includegraphics[width=0.45\textwidth]{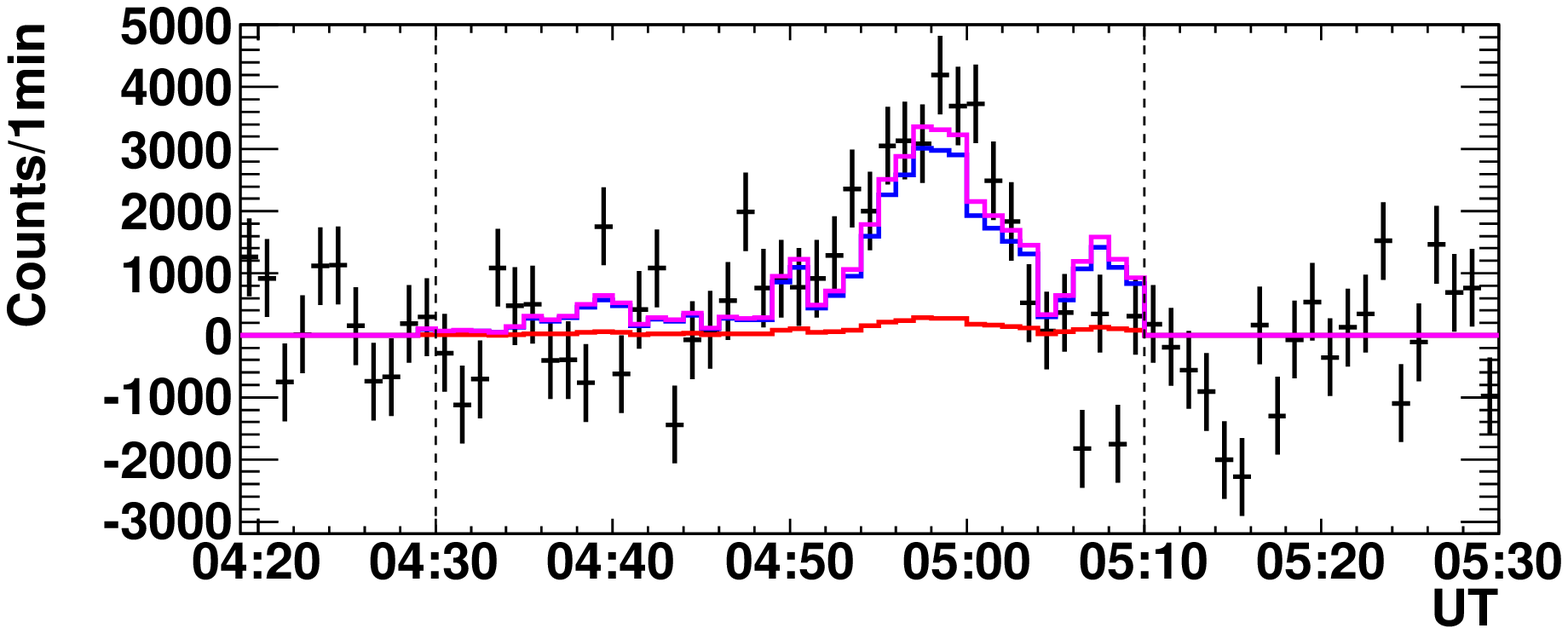}
\includegraphics[width=0.45\textwidth]{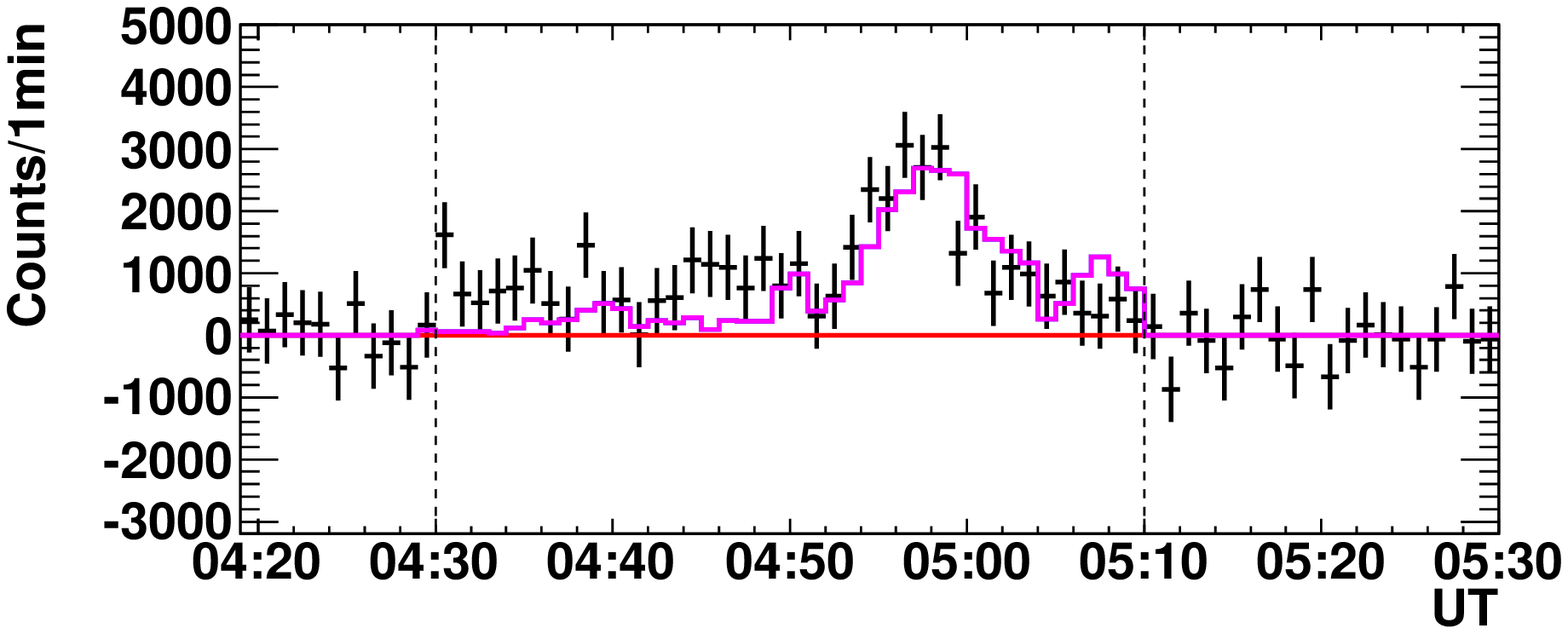}
\caption{Comparison of background-subtracted count histories with expected counts for 100722.
Left and right panels correspond to YBJ NM and SNT ($>$40 MeV) with 
anti-coincidence, respectively. Top and bottom panels represent the negative  and bipolar emissions, respectively.
Blue and red lines in each panel
indicate a count history predicted by $\gamma$ rays and neutrons, respectively. 
Each magenta line
shows an expected count history, which is summed over the counts from the relevant 
particles. Two vertical dashed lines in each panel denote the start and end times of
the count increase.
}
\label{fig:20100722_NMNT_TPcomp}
\end{figure}

The present simulation clearly showed that an NM64 neutron monitor, 
which was also used by \citet{armenia_2010}, 
has low, but not negligible, sensitivity to $\gamma$ rays. In addition, 
the survival probability of neutrons and $\gamma$ rays at the Aragats observatory would
not largely change from the present one (Fig.~\ref{fig:n_rate_spe}),  
because the air density at the Aragats observatory, which is $\sim$$9 \times 10^{-4}\,\, \mathrm{g\,cm^{-3}}$, is not 
very different from that  at the Yangbajing site, which is $\sim$$8\times 10^{-4}\,\, \mathrm{g\,cm^{-3}}$.
In fact, using the GEANT4 simulation, \citet{armenia_2010} derived $2.3\times10^{-3}$ as survival probability of neutrons arriving at their 
observatory, assuming bremsstrahlung $\gamma$ rays propagate
over 1500 m. This value is nearly consistent with the survival probability of neutrons that 
is derived in the present study for $H=1500$ m, which is $5\times10^{-4}$$-$$2\times10^{-3}$ (Fig.~\ref{fig:n_rate_spe}). 
Consequently, not neutrons but $\gamma$ rays may possibly dominate enhancements detected by 
the Aragats neutron monitor.

\subsubsection{Number of neutrons produced}%
Using the derived value of $\alpha_\mathrm{w}$, we 
evaluate the fluence of neutrons, $f_\mathrm{n}$, arriving at the observatory in energies  1 keV$-$300 MeV,  
by the following formula:
\[
f_\mathrm{n} = \alpha_\mathrm{w} \Delta T \int_{\Delta T} I(t) dt\int_{1\,\mathrm{keV}}^{300\, \mathrm{MeV}} 
S_\mathrm{n}(E_\mathrm{n})dE_\mathrm{n} = 1.4 \times 10^4 \,\, \mathrm{m^{-2}}, 
\]
where $\Delta T = 2400$ s and $S_\mathrm{n}(E_\mathrm{n})$ represents the simulated neutron spectrum,
assuming $\beta$ and $H$ are $-2$ and 900 m (Table~\ref{tab:chi2_alpha}), respectively. 
\citet{Carlson_Npro_2010} and \citet{Babich_prediction_gn_2010} described that photonuclear
neutrons produced by energetic $\gamma$ rays are observable at ground level when a $\gamma$-ray source 
is locate $<$5 km, since the neutron fluence is expected as (0.03$-$1)$\times10^4$ $\mathrm{m^{-2}}$ for the
former prediction and $10^3$$-$$10^7$ $\mathrm{m^{-2}}$   for the latter one. Actually, the value of 
$f_\mathrm{n}$ is consistent with their predictions. Thus, 
this agreement imply that the photonuclear reaction certainly occurs during mature stages of thunderclouds.

\section{Summary}
The prolonged $\gamma$-ray event, lasting for 40 min,  was observed on 2010 July 22 at Yangbajing in Tibet, China. 
Such a long-duration event associated with thunderstorms have never been observed. 
In addition, the present observations clearly showed that $\gamma$ ray extending to energies $>$40 MeV
were detected by SNT and very likely by YBJ NM.  Given these results, the present emissions strongly suggest that electrons are accelerated 
beyond at least 40 MeV in 40 min, by quasi-stable electric fields, which were formed 
during the mature stage of summer thunderclouds. The present duration is at least 5 times longer than those observed in
winter thunderstorms at the coastal area of the Japan sea. Probably, one of the main reasons for this difference 
would be ascribed to a difference in life cycles of mature stages of winter and summer thunderclouds.

The high-energy $\gamma$ rays would produce neutrons
via the photonuclear reaction of $^{14}$N($\gamma, n){}^{13}$N. 
The present simulation showed that the arriving neutron flux at $>$1 keV is 
expected to be lower than that of arriving $\gamma$ rays at $>$10 MeV by more than two orders of 
magnitude. Moreover, it revealed that unlike previously believed, neutron monitors 
are not insensitive to $\gamma$ rays. Consequently, it is found that 
bremsstrahlung $\gamma$ rays largely attribute the signal obtained by YBJ NM and photonuclear neutrons give only a small contribution to the signal.
The present study demonstrated that world-wide networks of neutron monitors~\cite{NMnetwork} and solar neutron telescopes~\cite{SNT_2001,Ymatsu_ICRC2007}
are useful for observations of  thunderstorm-related $\gamma$-ray emissions.   

\begin{acknowledgments}
The  study  is supported in parts by a Grant-in-Aid for Scientific Research (c), No. 20540298
and a Grant-in-Aid for Young Scientists (B), No. 19740167.  
This study is also supported in parts by the Special Posdoctoral Research Project for Basic Science in RIKEN; the 
Special Research Project for Basic Science in RIKEN ("Investigation of Spontaneously Evolving Systems"). 
\end{acknowledgments}

\newpage 

\clearpage
%
%
\begin{table}
\caption{\label{tab:increase_two_events}
Net count increases and statistical significance.}
\begin{ruledtabular}
\begin{tabular}{lllc}
{}                                  & $\Delta$ N\footnotemark[1] (significance) \\ \hline
YBJ NM                        & 34000 $\pm$ 4200 ($8.1\sigma$)               \\
SNT $>$40 MeV          & 44000 $\pm$ 3500 ($13\sigma$)                \\
SNT $>$80 MeV          & 16000 $\pm$ 2400 ($6.7\sigma$)               \\
SNT $>$120 MeV        & 8700 $\pm$ 1500 ($5.8\sigma$)                 \\
SNT $>$160 MeV        & 4600 $\pm$  970 ($4.7\sigma$)                  \\
\end{tabular}
\end{ruledtabular}
\footnotetext[1]{Each quoted error includes fluctuations of the background and total observed counts.}
\end{table}
\clearpage
%
%
\begin{table}
\caption{\label{tab:chi2_alpha}
$\chi^2$ minima and spectrum parameters determined.}
\begin{ruledtabular}
\begin{tabular}{cccccc}
{}                   &\multicolumn{2}{c}{Negative emission} & \multicolumn{2}{c}{Bipolar emission}   \\ \hline
{}                   & YBJ NM      & SNT                                & YBJ NM             & SNT   \\
 $\chi^2/d.o.f.$ & 49.4/39       & 46.2/39                          & 110/39            & 59.7/39      \\
$\alpha$ $(\times10^3\,\,\mathrm{MeV^{-1}m^{-2}s^{-1})}$\footnotemark[1]
	& $4.3\pm0.3$  & $4.2\pm0.3$  & $2.7 \pm 0.2$ & $2.8 \pm 0.2$   \\
 ($\beta$\footnotemark[2], H\footnotemark[3])  &\multicolumn{2}{c}{$(-2, 0.9)$}  & \multicolumn{2}{c}{$(-2, 0.6)$}    \\
\end{tabular}
\end{ruledtabular}
\footnotetext[1]{A normalization factor of an assumed power-law gamma-ray spectrum.} 
\footnotetext[2]{An estimated photon index of a power-law gamma-ray spectrum.} 
\footnotetext[3]{A source height (km) estimated.} 
\end{table}
\clearpage
%
%
\clearpage
\clearpage
%
%
%
\clearpage
%
%
\clearpage
%
%
%
%
\clearpage

%
%
\clearpage
%
%
\clearpage
%
%
\clearpage
%
%
\clearpage
%
%
\clearpage
%
%
\clearpage

\end{document}